\documentclass[aps,prl,twocolumn,10pt,amsmath,amssymb,bibnotes,superscriptaddress,longbibliography]{revtex4-1}

\usepackage{graphicx, braket}
\usepackage{float}
\usepackage{siunitx}
\usepackage[colorlinks,urlcolor=blue,citecolor=blue,linkcolor=blue]{hyperref}
\newcommand{\is}{\ensuremath{I_{CS}}}
\newcommand{\vp}{\ensuremath{V_D}}

\newcommand{\di}{\ensuremath{{I'_{CS}}}}

\newcommand{\tk}{\ensuremath{T_K}}
\usepackage{xcolor}

\usepackage[normalem]{ulem}
\newsavebox{\panelA}\newsavebox{\panelB}

\begin{document}

\title{Entropic signatures of the single-impurity Kondo state}

\author{Johann Drayne}
	\affiliation{Stewart Blusson Quantum Matter Institute, University of British Columbia, Vancouver, British Columbia, V6T1Z4, Canada}
	\affiliation{Department of Physics and Astronomy, University of British Columbia, Vancouver, British Columbia, V6T1Z1, Canada}
\author{Silvia L\"{u}scher}
\email{luescher@physics.ubc.ca}
	\affiliation{Stewart Blusson Quantum Matter Institute, University of British Columbia, Vancouver, British Columbia, V6T1Z4, Canada}
	\affiliation{Department of Physics and Astronomy, University of British Columbia, Vancouver, British Columbia, V6T1Z1, Canada}
\author{Will Grant}
	\affiliation{Stewart Blusson Quantum Matter Institute, University of British Columbia, Vancouver, British Columbia, V6T1Z4, Canada}
	\affiliation{Department of Physics and Astronomy, University of British Columbia, Vancouver, British Columbia, V6T1Z1, Canada}
\author{Vahid Movahed}
	\affiliation{Stewart Blusson Quantum Matter Institute, University of British Columbia, Vancouver, British Columbia, V6T1Z4, Canada}
	\affiliation{Department of Physics and Astronomy, University of British Columbia, Vancouver, British Columbia, V6T1Z1, Canada}
\author{Tim Child}
	\affiliation{Stewart Blusson Quantum Matter Institute, University of British Columbia, Vancouver, British Columbia, V6T1Z4, Canada}
	\affiliation{Department of Physics and Astronomy, University of British Columbia, Vancouver, British Columbia, V6T1Z1, Canada}
\author{Saeed Fallahi}
	\affiliation{Department of Physics and Astronomy, Purdue University, West Lafayette, Indiana, USA}
\author{Geoffrey C. Gardner}
	\affiliation{Microsoft Quantum, West Lafayette, Indiana, USA}
\author{Michael J. Manfra}
	\affiliation{Department of Physics and Astronomy, Purdue University, West Lafayette, Indiana, USA}
    \affiliation{Microsoft Quantum, West Lafayette, Indiana, USA}
   \affiliation{Elmore Family School of Electrical and Computer Engineering, Purdue University, West Lafayette, Indiana, USA}
   \affiliation{School of Materials Engineering, Purdue University, West Lafayette, Indiana, USA}
   \affiliation{Purdue Quantum Science and Engineering Institute, Purdue University, West Lafayette, Indiana, USA}

\author{Yaakov Kleeorin}
	\affiliation{Center  for  the  Physics  of  Evolving  Systems,  University  of  Chicago  ,  Chicago,  IL,  60637,  USA}
\author{Yigal Meir}
	\affiliation{Department of Physics, Ben-Gurion University of the Negev, Beer Sheva 84105, Israel}
\author{Joshua Folk}
\email{jfolk@physics.ubc.ca}
	\affiliation{Stewart Blusson Quantum Matter Institute, University of British Columbia, Vancouver, British Columbia, V6T1Z4, Canada}
	\affiliation{Department of Physics and Astronomy, University of British Columbia, Vancouver, British Columbia, V6T1Z1, Canada}
\date{\today}

\begin{abstract}
The Kondo singlet---a many-body state formed by entanglement between a localized spin and the Fermi sea---has been studied extensively through its transport signatures in quantum dots.  Here we report a thermodynamic measurement of the entropy suppression associated with the formation of the Kondo singlet, using temperature-dependent charge sensing and a Maxwell relation to track the suppression of spin entropy as the first electron is added to a strongly-coupled GaAs quantum dot.  Plotting $dN/dT$ against the simultaneously measured occupation $N$ reveals an asymmetric lineshape with its peak shifted to $N>1/2$---a hallmark of Kondo screening---that weakens with increasing temperature and is qualitatively reproduced by numerical renormalization group (NRG) calculations, with a small but persistent offset to lower occupation relative to the theory.  An independent measurement of conductance versus occupation on the same device provides a test of these quantities through the mixed-valence crossover and matches NRG within experimental uncertainty.
\end{abstract}
\maketitle

The Kondo effect is the textbook example of how exchange coupling between a localized spin and a metallic reservoir can reorganize the low-energy physics of a macroscopic system \cite{Kondo.1964, Hewson.1993}.  Below a characteristic scale $\tk$, the local moment is screened by a cloud of conduction electrons to form a many-body singlet whose defining feature is the entanglement between the impurity spin and the Fermi sea \cite{Nozieres.1974, mora2015fermi}.  The effect was first identified through the resistivity anomaly in bulk metals with dilute magnetic impurities~\cite{DeHaas.1934}, and later explained by Kondo's 1964 calculation.  More direct evidence for the singlet came nearly a decade later from heat-capacity experiments on dilute magnetic alloys, which revealed the vanishing entropy of the screened state \cite{gruner1978low,triplett1971calorimetric}.

Semiconductor quantum dots offer a controllable platform for studying the Kondo effect one impurity at a time, with the hybridization, and hence $\tk$, tunable in situ.  Single-impurity Kondo physics in quantum dots has been explored extensively through transport measurements~\cite{Goldhaber-Gordon.1998, Cronenwett.1998, Pustilnik.2004, vanderWiel.2000}. Yet transport accesses the Kondo resonance in the density of states, not the thermodynamic response of the screened state.  A thermodynamic signature of screening---the saturation of the impurity susceptibility at low temperature---has been observed in the context of the charge Kondo effect~\cite{piquard2023observing}, but the entropy of the spin-Kondo ground state has not yet been measured directly. A direct measurement of entropy provides an unambiguous signature of ground-state degeneracy and many-body entanglement that cannot be extracted from conductance alone.

Direct entropy measurement is difficult at the scale of a single impurity because the expected signal is of order $k_B$.  Recent experiments have nevertheless demonstrated that the entropy change associated with adding a single electron to a quantum dot can be extracted by measuring the temperature dependence of the dot occupation, $dN/dT$, and applying the Maxwell relation $(\partial S/\partial \epsilon)_T = -(\partial N/\partial T)_\epsilon$ \cite{Hartman.2018, child2022entropy, Adam2025Entropy, Pyurbeeva.2021x0i}.  Measuring the entropy associated with the formation of the Kondo singlet is especially challenging because the state forms when the dot is strongly coupled to the reservoir~\cite{child2022entropy}.  Strong hybridization broadens the charge transition over a wide range of dot energy, $\epsilon$, making the occupation nearly temperature-independent at any given $\epsilon$ and suppressing the measured signal $dN/dT$ to a small fraction of its weakly-coupled value.  An earlier measurement~\cite{child2022entropy} from our group did not resolve the entropy suppression associated with the Kondo singlet, limited by less effective averaging and, most importantly, by lack of investigation at lower temperatures.

Here we show that the suppression of spin entropy associated with the formation of the Kondo singlet can be robustly measured, using an optimized protocol to extract device parameters and to separate the small entropy signal from noise at electron temperatures of $30\,\mathrm{mK}$ and below.  The measured quantity is $dN/dT$ as a function of the dot occupation $N$ across the $N=0\rightarrow 1$ transition; by the Maxwell relation, this reflects how rapidly the entropy changes as the dot level is brought below the chemical potential of the leads.  In the absence of a Kondo effect, $dN/dT$ peaks at or even below $N=1/2$~\cite{child2022entropy}. With Kondo screening, however, the spin entropy remains quenched as the electron enters, and is released only deeper in the transition---where the Kondo temperature $\tk$ falls below $T$ and the local moment is restored---shifting the peak to $N>1/2$.
A measurement of conductance as a function of occupation in the same device, under similar conditions, is in quantitative agreement with numerical renormalization group (NRG) calculations, providing an independent confirmation of the Kondo state.

\begin{figure}
   \centering
   \includegraphics[width=1.0\columnwidth]{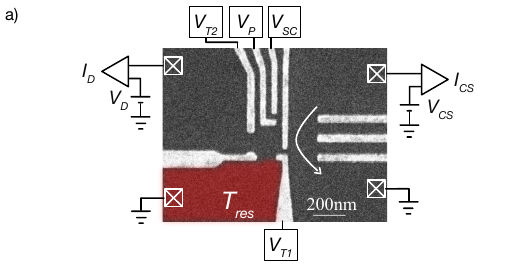}\\[2pt]
   \includegraphics[width=1.0\columnwidth]{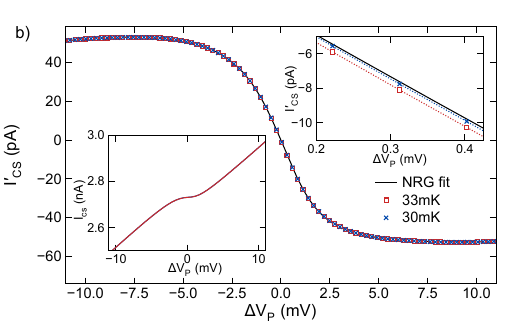}
   \caption{\label{fig:exptoverview}
   (a) Scanning electron micrograph of the device. The quantum dot (center) is coupled to a reservoir (red) whose temperature $T_\mathrm{res}$ is controlled by Joule heating; a second reservoir (upper left) can be opened for conductance measurements. A charge sensor (right) detects dot occupation. Cross-capacitance between $V_P$ and the charge sensor was reduced using the screening gate $V_{SC}$, \cite{supplement}.  All gates were depleted in the experiment, but on the labelled gates are referred to in the text.  (b) Charge-sensor current after background subtraction, $I'_{CS}$, across the $N=0\to 1$ transition, at $T=30$ and 33~mK. Left inset: raw \is~before subtraction. Right inset: zoom-in near $\Delta V_P=0.3~\mathrm{mV}$ showing the offset between hot and cold traces that reflects $dN/dT$; the solid black curve is a fit to the NRG occupation lineshape, from which the occupation N is extracted. Data collected with 210~$\mu \mathrm{V}$ applied to the charge sensor.
   }
\end{figure}

Measurements were performed on a quantum dot defined electrostatically in a GaAs heterostructure, next to a quantum point contact charge sensor to read out the dot occupation (Fig.~\ref{fig:exptoverview}a).  For entropy measurements, the dot is tunnel-coupled to a single reservoir whose temperature, $T_\mathrm{res}$, is controlled by Joule heating~\cite{child2022robust}, with $\Delta T/T\sim 0.1$.  Coupling to a second reservoir could be opened for conductance measurements.  Tunnel barriers to the two reservoirs were tuned by $V_{T1},V_{T2}$ respectively, controlling the hybridization $\Gamma$ of the dot with its leads.  Throughout this text, we refer to $\Gamma$ as an effective temperature in units of mK.  Experiments were carried out in a dilution refrigerator; all temperatures reported here are reservoir electron temperatures, measured directly from the width of charge transitions onto the dot at weak coupling \cite{supplement}, with an uncertainty of 2-3~mK.

Figure~\ref{fig:exptoverview}b shows charge-sensor traces across the $0\to 1$ transition at two reservoir temperatures, $T$ and $T+\Delta T$, where $\Delta T$ is set by Joule heating~\cite{child2022entropy, child2022robust}.  The dot energy, $\epsilon$, is controlled by a plunger gate voltage $V_P$; in what follows we report $\Delta V_P \equiv V_P - V_P(N=1/2)$.  Depending on the measurement, $V_P$ denotes either the bare plunger gate (entropy measurement) or a virtual gate that sweeps the plunger and screening gates together to cancel the plunger's cross-capacitance to the charge sensor (conductance measurement); in both cases we report $\Delta V_P$, and refer to \cite{supplement} for details.  Raw data are shown in the left inset, while the main panel shows the transition with a linear background subtracted.  At strong coupling, the occupation curve is broadened primarily by $\Gamma$ rather than by temperature, so the difference between the two curves, $\Delta I_{CS} \equiv I_{CS}(T+\Delta T)-I_{CS}(T)$, reflects a change in occupation of only $3\times 10^{-3}\,e$ or less between hot and cold data (Fig.~\ref{fig:exptoverview}b right inset), and requires extensive averaging to pull out of experimental noise. The extraction of such a small response to temperature change is made possible by rapid modulation of the reservoir temperature (36~Hz), enabling what is effectively a lockin measurement of $\Delta I_{CS}$, which itself is proportional to $dN/dT$---the raw input to the Maxwell relation.

\begin{figure}
   \includegraphics[width=1.0\columnwidth]{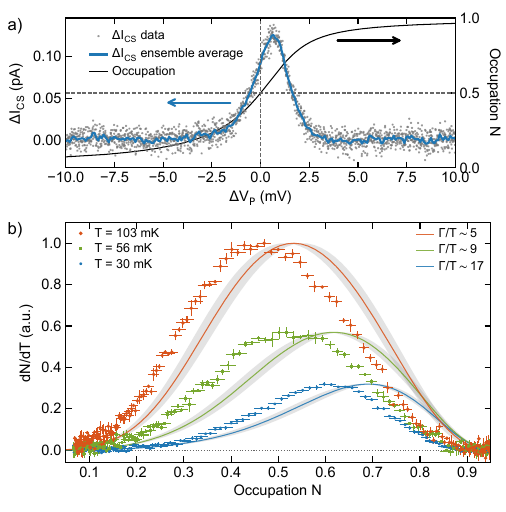}
   \caption{\label{fig:dndtvsn}
   (a) Difference of the 30 and 33~mK traces in Fig.~\ref{fig:exptoverview}b, $\Delta I_{CS} \propto dN/dT(\Delta V_P)$ (blue, ensemble average over 8 runs; gray markers: individual runs), together with the occupation $N(\Delta V_P)$ extracted from an NRG fit to the charge-sensor signal (black). The dashed lines marks $N=1/2$; the shift of the $dN/dT$ peak to $N>1/2$ is the central result of this paper.
   (b) $dN/dT$ (arb.\ units) plotted against $N$ for $T_\mathrm{res}=$30, 56, and 103~mK (markers), with NRG calculations (solid lines, labeled by $\Gamma/T$) for the same $\Gamma$ and $T$; all curves normalized to the peak of the 103~mK NRG curve. Shading around NRG curves represents experimental uncertainty in determining $\Gamma$.  Error bars represent standard error of binned data from multiple runs, and uncertainty in determining $N$ from $V_P$ \cite{supplement}.  Data collected with 210~$\mu V$ applied to the charge sensor.
   }
\end{figure}

The central comparison of this experiment is between the onset of spin entropy and the addition of charge, as the dot level is pulled below the Fermi energy of the leads.  Through the Maxwell relation, $dN/dT$ maps to $dS/d\epsilon$, so the peak of $dN/dT$ marks the gate voltage at which entropy emerges most rapidly.  We display the entropy signal in two ways.  In Fig.~\ref{fig:dndtvsn}a, it is shown in real units as the raw difference $\Delta I_{CS}(\Delta V_P)$, together with the simultaneously measured occupation $N(\Delta V_P)$. Extracting $N$ from the charge-sensor signal requires identifying scale factors, backgrounds and offsets; all of these are obtained from an NRG fit \cite{supplement}.  In Fig.~\ref{fig:dndtvsn}b, a normalized $dN/dT$ is instead plotted directly against $N$, because the lineshape of $dN/dT(N)$ shows most clearly the degree of Kondo screening.

Already in Fig.~\ref{fig:dndtvsn}a, the peak of $dN/dT$ sits past the midpoint toward $N>1/2$, though the offset is small compared with the width of the peak and easy to overlook on the $\Delta V_P$ axis.  In Fig.~\ref{fig:dndtvsn}b this shift is unmistakable: $dN/dT(N)$ is strongly asymmetric in the low temperature data, with its peak well to the right of $N=1/2$.  The asymmetry reflects how the Kondo temperature varies across the transition: $\tk$ is of order $\Gamma$ near half-occupation but falls exponentially as the level drops below the leads chemical potential.  When $\tk\gg T$, near $N=1/2$, the spin is screened into the singlet and its entropy suppressed; once $\tk$ drops below $T$, deeper in the transition, the screening fails: the local moment, and its entropy, are restored.  Because $\tk$ crosses $T$ on the occupied side of the transition, the spin entropy emerges there, pulling the peak of $dN/dT(N)$ toward $N>1/2$.

Figure~\ref{fig:dndtvsn}b also shows how this asymmetry evolves with temperature.  As $T$ increases the strong-coupling shift toward $N>1/2$ recedes and the peak of $dN/dT(N)$ moves back toward $N=1/2$.  This follows the same picture: at higher $T$, $\tk$ falls below $T$ when the dot level is closer to the lead chemical potential, that is, when $N$ is closer to $1/2$, so the point at which the local moment is restored moves toward half-occupation.  We note that this is not an approach to a symmetric peak: in the weak-coupling limit, $T\gg\Gamma$, the peak crosses to $N<1/2$ \cite{Hartman.2018,child2022entropy}.

To compare with NRG, we extract $\Gamma$ for the tuning of Fig.~\ref{fig:dndtvsn} from the temperature dependence of the charge transitions, measured from 25 to 800~mK \cite{supplement}.  NRG calculations at the corresponding $\Gamma/T$ reproduce the trend qualitatively, but quantitatively the measured asymmetry is weaker than predicted: the experimental $dN/dT(N)$ curves sit to the left of the NRG curves at all temperatures.  Although the mismatch in Fig.~\ref{fig:dndtvsn}b is small, it was consistent over multiple cooldowns of two different devices, and did not depend on the specific gate-voltage tuning of the dot.

For further insight into this persistent discrepancy with NRG, we performed an independent measurement on the same dot, probing the mixed-valence Kondo state through its conductance rather than its entropy.  Opening the pinched-off lead to a second reservoir and tuning the barriers to symmetric tunnel coupling, we measured the linear-response conductance $G$ and the charge-sensor occupation $N$ simultaneously as a function of $\Delta{V}_P$ \cite{supplement}, so that $G$ could be plotted directly against $N$---in the same parametric representation used for $dN/dT(N)$---and compared to NRG without any shifting or rescaling of the gate-voltage axis.

\begin{figure}
   \includegraphics[width=1.0\columnwidth]{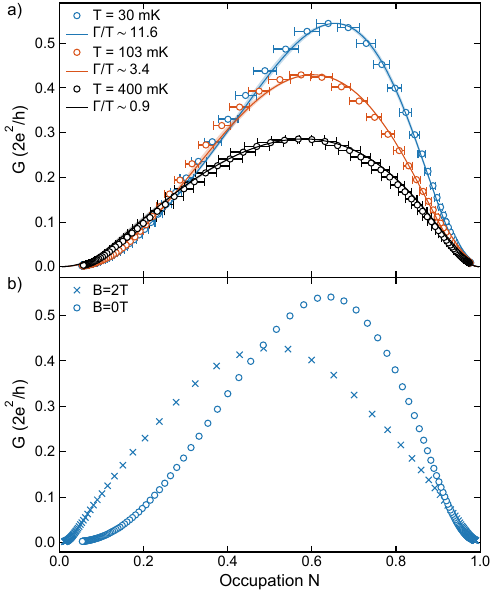}
   \caption{\label{fig:gvsn}
   (a) Conductance $G=dI_D/dV_D$ plotted against the simultaneously measured occupation $N$ across the $0\to 1$ transition of the same dot used in Figs.~\ref{fig:exptoverview} and \ref{fig:dndtvsn}, for three different temperatures. Markers: data; solid lines: NRG. Shaded bands around the NRG describe experimental uncertainty in determining $\Gamma$; error bars reflect uncertainty in determining $N$ from ${V}_P$ \cite{supplement}.
   (b) $G(N)$ at $B_\parallel=0$ (the 30mK data from panel a) and $B_\parallel=2$~T.  At high field the asymmetry of $G(N)$ collapses and the peak recovers a nearly symmetric shape, confirming that the shift in (a) originates in Kondo correlations rather than a device artifact.  Data collected with 50~$\mu V$ applied to the charge sensor.
   }
\end{figure}

As for $dN/dT$, NRG predicts the Kondo-screened conductance peak to shift towards $N>1/2$, reaching the unitary limit midway through the $N=1$ Coulomb valley for a fully-developed resonance.  The measured $G(N)$ reproduces this shift and, most importantly, matches NRG quantitatively across the entire transition (Fig.~\ref{fig:gvsn}a).  The Kondo origin of the asymmetry is confirmed independently by its in-plane field dependence: a Zeeman splitting $g\mu_B B_\parallel$ exceeding $k_B\tk$ prevents the singlet from forming, similar to the effect of raising the temperature above $\tk$~\cite{Costi2000Kondo,Cronenwett.1998,Goldhaber-Gordon.1998}. At $B_\parallel=2$~T ($g\mu_B B_\parallel \approx 50~\mu\mathrm{eV}$), the shift collapses and $G(N)$ recovers a nearly symmetric shape (Fig.~\ref{fig:gvsn}b). Together, the parameter-free agreement with NRG and the field-induced collapse establish that the conductance asymmetry is a fingerprint of the Kondo singlet.  Figure~\ref{fig:gvsn}a is, to our knowledge, the first simultaneous measurement of $G$ and $N$ through the mixed-valence crossover. The close agreement with theory is, itself, a non-trivial validation of NRG in a regime where analytical methods break down.

This same agreement strongly constrains the possible explanations for the residual discrepancy in the entropy measurement (Fig.~\ref{fig:dndtvsn}b).  We consider three classes of explanation: (i)~the Kondo correlations in the device are genuinely weaker than in the modeled system; (ii)~NRG mis-computes $dN/dT(N)$ in the mixed-valence regime; or (iii)~the measured $dN/dT(N)$ does not faithfully represent the entropic signal---either because the occupation $N$ is mis-measured, distorting the horizontal axis of Fig.~\ref{fig:dndtvsn}b, or because the measured temperature dependence reflects charge beyond the Kondo-induced change in dot occupation.

The $G(N)$ comparison  bears on all three classes of explanation.  A genuinely weaker Kondo state---whether from charge-sensor back-action dephasing the singlet~\cite{silva2003peculiarities, avinun2004controlled, kang2005decoherence, kang2007entanglement}, a mis-estimated $\Gamma$, or an error in electron temperature---would equally suppress the Kondo signatures in the conductance, so the quantitative match strongly disfavors~(i).  Because $G(N)$ and $dN/dT(N)$ follow from the same NRG solution at the same parameters, the agreement of $G(N)$ makes it unlikely that the NRG prediction for $dN/dT(N)$ is itself wrong~(ii). Furthermore, because the comparison is made against the same $N$ used in Fig.~\ref{fig:dndtvsn}b, any distortion of $N$ would equally spoil the $G(N)$ match, all but excluding the first branch of~(iii).

This leaves the second branch of~(iii).  That the charge sensor responds to more than the dot occupation is not in itself the issue---it always does; the question is whether the temperature affects any charge registered by the charge sensor \emph{besides} the Kondo-induced occupation of the dot.  Any such change would add a non-entropic term to the measured $dN/dT$, displacing it from the entropic $dN_\mathrm{dot}/dT$ that the Maxwell relation and NRG describe.  For example, a temperature-dependent redistribution of charge in or near the dot---a shift in the screening cloud in the leads, or in the spatial extent of the dot wavefunction---would register on the sensor without changing the number of electrons on the dot.  Alternatively, a weak temperature dependence of the hybridization $\Gamma$, and with it $\tk$, would shift the dot occupation through the temperature dependence of the lineshape itself: a real $dN_\mathrm{dot}/dT$, but of hybridization rather than entropic origin, and therefore absent from the NRG prediction which assumes fixed $\Gamma$. Measuring $dN/dT(N)$ at high Zeeman energy, where the Kondo singlet is destroyed (Fig.~\ref{fig:gvsn}b), would isolate these background contributions directly. This was not feasible here, however: $dN/dT$ could no longer be resolved above the charge-sensor noise, which grew significantly at high field---likely from EMF induced in the measurement lines by cryostat vibrations.

In summary, we have observed the suppression of spin entropy associated with the formation of a Kondo singlet when the first electron enters a strongly-coupled quantum dot---to our knowledge, the first entropy determination of a state whose defining feature is the entanglement between an impurity spin and the Fermi sea.  The entropy signature---an asymmetry of $dN/dT(N)$ that emerges as $T$ drops below $\tk$---is reproduced qualitatively by NRG, while a simultaneous measurement of $G(N)$ matches NRG quantitatively, confirming that the model parameters for the Kondo state match the experiment.  A residual discrepancy between the measured and predicted $dN/dT(N)$ may reflect the sensitivity of the charge sensor to temperature-induced charge redistribution beyond the dot itself.

Looking forward, one would like to observe the complete suppression of spin entropy that marks a fully-formed Kondo singlet---a limit the present mixed-valence measurement never reaches.  Reaching it would require $\tk\gg T$ while keeping the charge transition sharp enough to resolve $dN/dT$; because increasing $\Gamma$ raises $\tk$ but also broadens the transition, this likely calls for decoupling the Kondo scale from the transition width, for instance by establishing Kondo screening through a confined reservoir.

ACKNOWLEDGEMENTS: This project has received funding from European Research Council (ERC) under the European Union's Horizon 2020 research and innovation program under grant agreement No 951541. Y. Meir acknowledges support by the Israel Science Foundation Breakthrough Fund (grant 737/24).  Experiments at UBC were undertaken with support from the Stewart Blusson Quantum Matter Institute, the Natural Sciences and Engineering Research Council of Canada, the Canada Foundation for Innovation, the Canadian Institute for Advanced Research, and the Canada First Research Excellence Fund, Quantum Materials and Future Technologies Program.  Work in the Manfra group at Purdue University was supported by the US DOE Office of Basic Energy Sciences under Award DE-SC0020138.

\bibliography{main}

@article{DeHaas.1934,
title = {The electrical resistance of gold, copper and lead at low temperatures},
journal = {Physica},
volume = {1},
number = {7},
pages = {1115-1124},
year = {1934},
issn = {0031-8914},
doi = {https://doi.org/10.1016/S0031-8914(34)80310-2},
url = {https://www.sciencedirect.com/science/article/pii/S0031891434803102},
author = {W.J. {de Haas} and J. {de Boer} and G.J. {van dën Berg}},
}

@article{Kondo.1964,
  author  = {Kondo, Jun},
  title   = {Resistance Minimum in Dilute Magnetic Alloys},
  journal = {Progress of Theoretical Physics},
  volume  = {32},
  number  = {1},
  pages   = {37--49},
  year    = {1964},
  doi     = {10.1143/PTP.32.37}
}

@book{Hewson.1993,
  author    = {Hewson, A. C.},
  title     = {The {Kondo} Problem to Heavy Fermions},
  publisher = {Cambridge University Press},
  year      = {1993},
  address   = {Cambridge, UK}
}

@article{Nozieres.1974,
  author  = {Nozi{\`e}res, P.},
  title   = {A ``{F}ermi-liquid'' description of the {Kondo} problem at
             low temperatures},
  journal = {Journal of Low Temperature Physics},
  volume  = {17},
  pages   = {31--42},
  year    = {1974},
  doi     = {10.1007/BF00654541}
}

@incollection{gruner1978low,
title = {Chapter 8 Low Temperature Properties of Kondo Alloys},
editor = {D.F. Brewer},
series = {Progress in Low Temperature Physics},
publisher = {Elsevier},
volume = {7},
pages = {591-647},
year = {1978},
issn = {0079-6417},
doi = {https://doi.org/10.1016/S0079-6417(08)60178-X},
url = {https://www.sciencedirect.com/science/article/pii/S007964170860178X},
author = {G. Grüner and A. Zawadowski},
}

@article{triplett1971calorimetric,
  title   = {Calorimetric Evidence for a Singlet Ground State in
             $\mathrm{Cu}\mathrm{Cr}$ and $\mathrm{Cu}\mathrm{Fe}$},
  author  = {Triplett, B. B. and Phillips, Norman E.},
  journal = {Physical Review Letters},
  volume  = {27},
  number  = {15},
  pages   = {1001--1004},
  year    = {1971},
  doi     = {10.1103/PhysRevLett.27.1001}
}

@article{piquard2023observing,
  title   = {Observing the universal screening of a {Kondo} impurity},
  author  = {Piquard, C. and Glidic, P. and Han, C. and Aassime, A.
             and Cavanna, A. and Gennser, U. and Meir, Y. and Sela, E.
             and Anthore, A. and Pierre, F.},
  journal = {Nature Communications},
  volume  = {14},
  number  = {1},
  pages   = {7263},
  year    = {2023},
  doi     = {10.1038/s41467-023-42817-6}
}

@article{Pustilnik.2004,
  author  = {Pustilnik, M. and Glazman, L.},
  title   = {Kondo effect in quantum dots},
  journal = {Journal of Physics: Condensed Matter},
  volume  = {16},
  pages   = {R513--R537},
  year    = {2004},
  doi     = {10.1088/0953-8984/16/16/R01}
}

@article{Goldhaber-Gordon.1998,
  author  = {Goldhaber-Gordon, D. and Shtrikman, Hadas and Mahalu, D.
             and Abusch-Magder, D. and Meirav, U. and Kastner, M. A.},
  title   = {Kondo effect in a single-electron transistor},
  journal = {Nature},
  volume  = {391},
  pages   = {156--159},
  year    = {1998},
  doi     = {10.1038/34373}
}

@article{Cronenwett.1998,
  author  = {Cronenwett, Sara M. and Oosterkamp, Tjerk H.
             and Kouwenhoven, Leo P.},
  title   = {A Tunable {Kondo} Effect in Quantum Dots},
  journal = {Science},
  volume  = {281},
  pages   = {540--544},
  year    = {1998},
  doi     = {10.1126/science.281.5376.540}
}

@article{vanderWiel.2000,
  author  = {van der Wiel, W. G. and De Franceschi, S. and Fujisawa, T.
             and Elzerman, J. M. and Tarucha, S. and Kouwenhoven, L. P.},
  title   = {The {Kondo} Effect in the Unitary Limit},
  journal = {Science},
  volume  = {289},
  pages   = {2105--2108},
  year    = {2000},
  doi     = {10.1126/science.289.5487.2105}
}

@article{Costi2000Kondo,
  author  = {Costi, T. A.},
  title   = {Kondo Effect in a Magnetic Field and the Magnetoresistivity
             of Kondo Alloys},
  journal = {Phys. Rev. Lett.},
  volume  = {85},
  pages   = {1504--1507},
  year    = {2000},
  doi     = {10.1103/PhysRevLett.85.1504}
}

@article{mora2015fermi,
  author  = {Mora, Christophe and Moca, C\u{a}t\u{a}lin Pa\c{s}cu
             and von Delft, Jan and Zar\'and, Gergely},
  title   = {{Fermi}-liquid theory for the single-impurity {Anderson} model},
  journal = {Physical Review B},
  volume  = {92},
  pages   = {075120},
  year    = {2015},
  doi     = {10.1103/PhysRevB.92.075120}
}

@article{Hartman.2018,
  author  = {Hartman, Nikolaus and Olsen, Christian and L{\"u}scher, Silvia
             and Samani, Mohammad and Fallahi, Saeed and Gardner, Geoffrey C.
             and Manfra, Michael and Folk, Joshua},
  title   = {Direct entropy measurement in a mesoscopic quantum system},
  journal = {Nature Physics},
  volume  = {14},
  pages   = {1083--1086},
  year    = {2018},
  doi     = {10.1038/s41567-018-0250-5}
}

@article{child2022entropy,
  author  = {Child, Timothy and Sheekey, Owen and L{\"u}scher, Silvia
             and Fallahi, Saeed and Gardner, Geoffrey C. and Manfra, Michael
             and Mitchell, Andrew and Sela, Eran and Kleeorin, Yaakov
             and Meir, Yigal and Folk, Joshua},
  title   = {Entropy measurement of a strongly coupled quantum dot},
  journal = {Physical Review Letters},
  volume  = {129},
  pages   = {227702},
  year    = {2022},
  doi     = {10.1103/PhysRevLett.129.227702}
}

@article{child2022robust,
  author  = {Child, Timothy and Sheekey, Owen and L{\"u}scher, Silvia
             and Fallahi, Saeed and Gardner, Geoffrey C and Manfra, Michael
             and Folk, Joshua},
  title   = {A robust protocol for entropy measurement in mesoscopic
             circuits},
  journal = {Entropy},
  volume  = {24},
  number  = {3},
  pages   = {417},
  year    = {2022},
  doi     = {10.3390/e24030417}
}

@article{Pyurbeeva.2021x0i,
  author  = {Pyurbeeva, Evgeniya and Mol, Jan A.},
  title   = {A Thermodynamic Approach to Measuring Entropy in a
             Few-Electron Nanodevice},
  journal = {Entropy},
  volume  = {23},
  pages   = {640},
  year    = {2021},
  doi     = {10.3390/e23060640}
}

@article{silva2003peculiarities,
  author  = {Silva, A. and Levit, S.},
  title   = {Peculiarities of the controlled dephasing of a quantum dot
             in the {Kondo} regime},
  journal = {Europhysics Letters},
  volume  = {62},
  number  = {1},
  pages   = {103},
  year    = {2003},
  doi     = {10.1209/epl/i2003-00368-1}
}

@article{avinun2004controlled,
  author  = {Avinun-Kalish, M. and Heiblum, M. and Silva, A.
             and Mahalu, D. and Umansky, V.},
  title   = {Controlled Dephasing of a Quantum Dot in the {Kondo} Regime},
  journal = {Physical Review Letters},
  volume  = {92},
  pages   = {156801},
  year    = {2004},
  doi     = {10.1103/PhysRevLett.92.156801}
}

@article{kang2005decoherence,
  title   = {Decoherence of the {Kondo} singlet via a quantum point contact detector},
  author  = {Kang, Kicheon},
  journal = {Physical Review Letters},
  volume  = {95},
  number  = {20},
  pages   = {206808},
  year    = {2005},
  publisher = {APS},
  doi     = {10.1103/PhysRevLett.95.206808}
}

@article{kang2007entanglement,
  title   = {Entanglement, measurement, and conditional evolution of the {Kondo} singlet interacting with a mesoscopic detector},
  author  = {Kang, Kicheon and Khym, Gyong Luck},
  journal = {New Journal of Physics},
  volume  = {9},
  number  = {5},
  pages   = {121--121},
  year    = {2007}
}

@article{Adam2025Entropy,
  author  = {Adam, C. and Duprez, H. and Lehmann, N. and Yglesias, A.
             and Denisov, A. O. and Cances, S. and Ruckriegel, M. J.
             and Masseroni, M. and Tong, C. and Huang, W. and Kealhofer, D.
             and Garreis, R. and Watanabe, K. and Taniguchi, T.
             and Ensslin, K. and Ihn, T.},
  title   = {Entropy Spectroscopy of a Bilayer Graphene Quantum Dot},
  journal = {Physical Review Letters},
  volume  = {135},
  number  = {12},
  pages   = {126202},
  year    = {2025},
  doi     = {10.1103/vbbj-138r}
}

@misc{supplement,
  note = {See Supplemental Material for details of the device, the virtual-gate
          construction, calibration procedures, NRG parameter extraction, and
          charge-sensor back-action checks.}
}

\clearpage
\onecolumngrid
\raggedbottom
\setcounter{figure}{0}
\renewcommand{\thefigure}{S\arabic{figure}}
\setcounter{equation}{0}
\renewcommand{\theequation}{S\arabic{equation}}
\setcounter{secnumdepth}{3}
\setcounter{section}{0}
\renewcommand{\thesection}{S\arabic{section}}
\renewcommand{\thesubsection}{S\arabic{section}.\arabic{subsection}}
\renewcommand{\thesubsubsection}{S\arabic{section}.\arabic{subsection}.\arabic{subsubsection}}

\makeatletter
\renewcommand\section{\@startsection{section}{1}{\z@}%
  {-3.5ex \@plus -1ex \@minus -.2ex}%
  {2.3ex \@plus.2ex}%
  {\raggedright\normalfont\Large\bfseries}}
\renewcommand\subsection{\@startsection{subsection}{2}{\z@}%
  {-3.25ex\@plus -1ex \@minus -.2ex}%
  {1.5ex \@plus .2ex}%
  {\raggedright\normalfont\large\bfseries}}
\renewcommand\subsubsection{\@startsection{subsubsection}{3}{\z@}%
  {-3.25ex\@plus -1ex \@minus -.2ex}%
  {1.5ex \@plus .2ex}%
  {\raggedright\normalfont\normalsize\bfseries}}
\def\p@subsection{}\def\p@subsubsection{}
\makeatother

\begin{center}
\Large\textbf{Supplementary Information:}\\Entropic signatures of the single-impurity Kondo state
\end{center}

\section{Devices and cooldowns}\label{sec:devcd}

The figures in the main text were all collected from a single device in a single cooldown, but we confirmed the results with another device measured over two other cooldowns.  We label the two devices MainDevice and SuppDevice, measured in cooldowns CD1, CD2, and CD3.

\begin{itemize}
\item MainDevice, CD1: Figures 1, 2, 3, S1, S2a, S3, S4, S5, S6, S7, S8, S12, S13
\item SuppDevice, CD2: Figures S9, S10
\item SuppDevice, CD3: Figure S11
\end{itemize}

The only design difference between MainDevice and SuppDevice was the lack of a screening gate, $V_{SC}$, in SuppDevice.

\section{Reservoir electron temperature calibration}\label{sec:Tcalib}

All temperatures quoted in the paper are reservoir electron temperatures $T_\mathrm{res}$ measured directly from charge-transition lineshapes at weak coupling to the reservoir, not mixing-chamber temperatures. When $\Gamma \ll k_B T$, the charge transition is thermally broadened, and the fitted width $\theta$ is linear in $T_\mathrm{res}$ as long as the reservoir tracks the mixing-chamber thermometer. At the lowest temperatures, $\theta$ saturates at a residual floor that is likely limited electronic noise propagating to the device or charge noise in the device.  This sets the effective lowest electron temperature accessible in the measurement.

\begin{figure}[H]
    \centering
    \includegraphics[width=0.4\columnwidth]{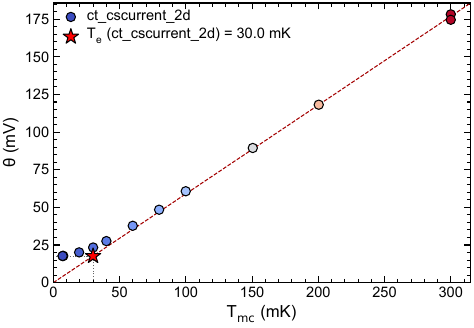}
    \caption{\label{fig:electrontemp}
    Charge-transition width vs.\ mixing-chamber temperature. Fitted charge-transition width $\theta$ plotted against $T_\mathrm{mc}$ for each dataset; marker colour encodes $T_\mathrm{mc}$. The dashed red line is the least-squares fit $\theta = m\,T_\mathrm{mc}$ over points with $T_\mathrm{mc} > 40\,\mathrm{mK}$. At the lowest $T_\mathrm{mc}$, $\theta$ saturates. The red star marks $T_{res,\mathrm{base}} \approx 30\,\mathrm{mK}$, obtained by projecting the saturated cold-point width back onto the linear fit (grey dotted guides).
    }
\end{figure}

We fit a straight line $\theta = m\,T_\mathrm{mc}$ to the subset of points with $T_\mathrm{mc} > 40\,\mathrm{mK}$, chosen just above the visible saturation knee. Writing $\theta_\mathrm{low}$ for the saturated charge-transition width at the coldest accessible $T_\mathrm{mc}$, the base reservoir electron temperature is $T_{res,\mathrm{base}} = \frac{\theta_\mathrm{low} - \theta_0}{m}$,
the temperature at which the linear extrapolation reproduces the cold-point width. The value of $T_{res,\mathrm{base}}$ varied from cooldown to cooldown, fridge to fridge, and from device to device.   Fig.~\ref{fig:electrontemp} shows the data for the cooldown from which the main text data was collected, giving $T_{res,\mathrm{base}} \approx 30\,\mathrm{mK}$ at the coldest mixing-chamber setting of $T_\mathrm{mc} \approx 7\,\mathrm{mK}$.  However, for some of the data used in the supplement we found $T_{res,\mathrm{base}} \approx 20\,\mathrm{mK}$.

\newpage
\section{Data acquisition}\label{sec:dataacq}

\subsection{Technical details}\label{sec:techdetails}

{\raggedright
Gate voltages and heater bias currents were set by DACs operating at a sampling rate of 9709~Hz. The charge sensor current was read out via a current-to-voltage converter using an ADC synchronized to the DAC steps. The synchronized DAC/ADC units were optimized by our group and the UBC Physics and Astronomy technical staff, starting from the OpenDAC platform (\url{https://opendacs.com/}) developed by Hugh Churchill and Andrea Young. Following the open-source approach from Young and Churchill, we are happy to share technical details with interested parties. Each gate is driven by two DAC channels with voltage dividers of 1:10 and 1:200 whose outputs are summed at the gate electrode, providing independent coarse and fine voltage control.
The prefix $\Delta$ in a gate-voltage label denotes sweeping the fine (1:200) channel of that gate while holding the coarse (1:10) channel fixed. To suppress the plunger's cross-capacitance to the charge sensor, the dot energy was in most cases swept with a virtual gate: the fine channels of the plunger gate and a compensating gate were stepped together in opposition, so that the compensating gate cancels the plunger-induced signal on the charge sensor. The main-device conductance measurement (Fig.~3) used the screening gate SC as the compensating gate, while the second-device entropy and conductance measurements (Figs.~\ref{fig:dev2_dndtvsn} and \ref{fig:dev2_gvsn}) used the charge-sensor gate. The exception is the main-device entropy measurement (Fig.~2), for which the plunger gate $V_P$ was swept alone, with the screening gate held at a fixed (mV-scale) potential. In all cases we denote the swept dot-energy axis $V_P$, so that $\Delta V_P$ refers to the dot energy throughout.  Comparing all virtual (and non-virtual) gate datasets, we found no difference in the final result due to the compensating gate.  The only effect was a more straightforward and robust data analysis procedure when an optimal compensating gate routine was applied.

\par}

\subsection{Entropy measurement}\label{sec:entmeas}

{\raggedright
For the entropy measurement the reservoir temperature was modulated by Joule heating using a periodic three-level square waveform applied simultaneously to the heater-bias resistors at the source and drain contacts, with opposite polarity so that the heated reservoir remained at zero potential. The heater cycle at each gate-voltage setting consisted of a repeated positive/zero/negative/zero sequence, producing a heat/cool/heat/cool pattern. The quoted heater step $dT = 3\,\mathrm{mK}$ refers to the reservoir-temperature increment between consecutive hot and cold segments, calibrated using the same approach as described in Section~\ref{sec:Tcalib}.
\par}

At each gate-voltage setting the heater cycle was completed before the gate voltage was stepped. Each hot or cold segment lasted 13.9~ms, corresponding to 135 DAC/ADC samples at 9709~Hz. When only the heater bias changed between segments, the first 6 samples were discarded to allow the circuit to settle. After each complete hot/cold/hot/cold cycle the gate voltage was stepped, and the first 11 samples of the new setting were discarded to remove transients associated with the gate-voltage step. This interlaced acquisition scheme is described in detail in Ref.~\cite{child2022robust}.

\subsection{Conductance measurement}\label{sec:condmeas}
{\raggedright
For the conductance measurement the second reservoir was opened so that the dot was symmetrically coupled to both leads. The device parameters $(\alpha, \Gamma)$ had to be redetermined for this configuration, as the gate settings differed from those of the entropy measurement; the measurement protocol and fitting procedure were slightly different from those in Sec.~\ref{sec:entanalysis} and are described in Sec.~\ref{sec:condctanalysis}.
\par}
A square-wave bias at $\approx 33\,\mathrm{Hz}$ with amplitude $V_\mathrm{bias} = 0.911\,\mu\mathrm{V}$ was generated by a DAC and applied to the source contact, and the resulting current was measured to extract the differential conductance $dI/dV$. The charge sensor, biased at $50\,\mu\mathrm{V}$, was read out simultaneously at each gate-voltage step.  The dot energy was swept using the combined plunger-- gate ${V}_P$ (see Sec.~\ref{sec:techdetails} for the gate construction). As a consequence, the  nearly linear background that dominates the raw traces in the entropy measurement was significantly reduced for the conductance measurement.
At each mixing-chamber temperature the dot energy sweep was repeated 50--100 times, each sweep taking $\approx 30\,\mathrm{s}$, with successive sweeps distributed over perpendicular fields $B_\perp = 40$--$60\,\mathrm{mT}$ so that averaging over repeats suppresses random noise and a weak quantum-interference artifact in the charge sensor, described in more detail in Sec.~\ref{sec:QI}. This produced a two-dimensional record of dot-conductance and charge-sensor currents versus gate voltage at each temperature.

\subsection{Charge sensor bias}\label{sec:csbias}

Faithful readout of the charge sensor is the single most important ingredient of this experiment, since the entire entropy determination rests on resolving a small, temperature-dependent change in the charge-sensor signal across the transition. The bias applied across the charge sensor sets the readout sensitivity, and its value is a compromise. Too little bias leaves the signal buried in noise, inflating the uncertainty in the occupation $N$ and in the entropy as functions of gate voltage. Too much bias raises the concern that the sensor perturbs the system it measures: a strongly biased charge sensor can dephase the Kondo singlet through back-action, as observed by Avinun-Kalish \emph{et al.}~\cite{avinun2004controlled} and analyzed theoretically in Refs.~\cite{silva2003peculiarities, kang2005decoherence, kang2007entanglement}.

\begin{figure}[H]
    \centering
    \begin{minipage}[t]{0.48\columnwidth}
        \textbf{(a)}\par\smallskip
        \includegraphics[width=\linewidth]{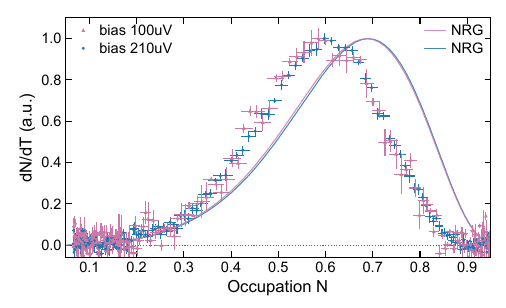}
    \end{minipage}\hfill
    \begin{minipage}[t]{0.48\columnwidth}
        \textbf{(b)}\par\smallskip
        \includegraphics[width=\linewidth]{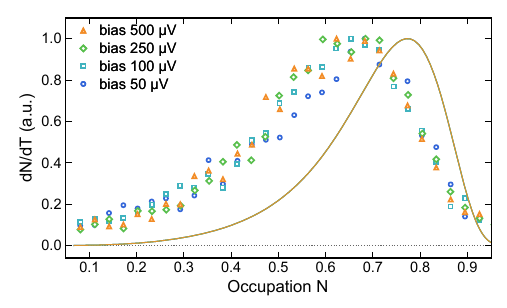}
    \end{minipage}
    \caption{\label{fig:bias_dep}
    Charge-sensor bias dependence of $dN/dT(N)$.
    \textbf{(a)} The Fig.~2 dataset measured at biases of $100$ and $210\,\mu\mathrm{V}$, with the NRG prediction. The two biases give indistinguishable lineshapes, confirming that the $210\,\mu\mathrm{V}$ bias used for Fig.~2 does not distort the measured entropy signature; the $100\,\mu\mathrm{V}$ data required substantially longer averaging to reach the same signal-to-noise.
    \textbf{(b)} The $\Gamma = 965\,\mathrm{mK}$ dataset (the setting of Sec.~\ref{sec:dev2_dndt}) at biases of $50$, $100$, $250$, and $500\,\mu\mathrm{V}$ (NRG shown for reference). No systematic dependence on bias is visible over this tenfold range, indicating that charge-sensor back-action does not affect the measurement at the biases used.
    }
\end{figure}

We therefore checked carefully that the bias we applied does not affect the result. The entropy measurement of Fig.~2 used a charge-sensor bias of $210\,\mu\mathrm{V}$. Repeating it at $100\,\mu\mathrm{V}$ gives an indistinguishable $dN/dT(N)$ (Fig.~\ref{fig:bias_dep}(a)), at the cost of much longer averaging to reach comparable signal-to-noise.

In Fig.~\ref{fig:bias_dep}(b) we confirm that this independence from charge-sensor bias extends over a much broader range, from $50$ to $500\,\mu\mathrm{V}$, in this case for the measurement setting later described in Sec.~\ref{sec:dev2_dndt}. The $dN/dT(N)$ lineshape shows no discernible dependence on bias across this tenfold range, beyond experimental uncertainty.

This is consistent with Avinun-Kalish \emph{et al.}~\cite{avinun2004controlled}, where detector-induced dephasing became appreciable only above $\sim 500\,\mu\mathrm{V}$. We stress, however, that the onset bias is not expected to be quantitatively the same from one device to another: the back-action dephasing rate scales as the square of the charge sensor's sensitivity to a charge transition on the dot, which depends on the geometry and tuning of the sensor and is therefore device specific.

\newpage
\section{Analysis of Entropy Data (Figs.~1 and 2)}\label{sec:entanalysis}

\subsection{Figure~2 details}\label{sec:fig2details}

\emph{Top panel.} Before averaging together the eight background-subtracted traces (Sec.~\ref{sec:traceproc}) represented in Fig.~\ref{fig:dndtvsn}a, they were aligned by first fitting to NRG then shifting the midpoints to zero.  After alignment, those raw data are plotted with grey markers.  The ensemble mean of the aligned traces is the blue curve. The occupation $N$ is the NRG occupancy evaluated at the best-fit $(\alpha, \Gamma)$, mapped onto the same aligned gate-voltage axis.

\emph{Bottom panel.} The measured $dN/dT(N)$ for $T_\mathrm{res} = 30$, 56, and 103~mK is shown together with the NRG predictions at $\Gamma_\mathrm{best}$ and shaded bands spanning $[\Gamma_\mathrm{lo}, \Gamma_\mathrm{hi}]$ (Sec.~\ref{sec:gammaunc}).
All data and theory curves were normalized by the peak of the 103~mK curve to place the three temperatures on a common vertical scale.\\

\noindent \textbf{Overview of analysis:} The raw data for the entropy measurement consisted of many repeated charge-sensor sweeps across the charge transition, recorded at each of several mixing-chamber temperatures. Extracting the $dN/dT$ curves shown in Fig.~2 required three stages of analysis.

The first stage is trace processing (Sec.~\ref{sec:traceproc}). At each temperature, individual sweeps (``traces'') were fit to a standard charge-transition model to extract their midpoints and widths. Traces whose width or fit residual lay far from the median were rejected as outliers. The surviving traces were shifted to align their midpoints, averaged to a single high-signal-to-noise trace, and then corrected for the nearly linear background that arises from cross-capacitance of the plunger gate to the charge sensor. The output of this stage is one clean, background-subtracted trace per temperature.

The second stage is a joint fit of the full temperature series  to numerical renormalization group (NRG) calculations (Sec.~\ref{sec:nrgproc}). Because the device parameters---the gate-voltage-to-energy rescaling (leverarm) $\alpha$ and the hybridization $\Gamma$---should not depend on temperature, they were treated as global parameters shared across all datasets.

The third stage is uncertainty propagation (Sec.~\ref{sec:gammaunc}). The shallow minimum of the constrained fit error $E(\Gamma)$ defines an interval $[\Gamma_\mathrm{lo},\Gamma_\mathrm{hi}]$ for $\Gamma$; evaluating the NRG prediction for $dN/dT(N)$ across this interval yields the shaded bands in Fig.~2(b).

\subsection{Trace processing}\label{sec:traceproc}

For each temperature point in the series, the same charge-transition sweep was repeated many times to build up statistics.  We refer to each sweep as a \emph{trace}; the collection of traces at one mixing-chamber temperature is a \emph{dataset} (indexed $i$); and the set of all datasets at different $T_\mathrm{mc}$ values is the \emph{temperature series}.

Each trace was fit individually to the model for a thermally broadened charge transition, as in Ref.~\cite{child2022robust}: a \emph{tanh} step of amplitude $I_1$ at midpoint $V_0$, thermally broadened with width $\theta$, sitting on a linear background of offset $I_0$ and slope $C_0$.
Then, traces were screened for outliers in the fitted width $\theta$ and the per-trace RMS residual $r$ based on their distance from the median.
To align the surviving traces, each was interpolated onto a common gate-voltage grid shifted by its fitted midpoint.  The aligned, outlier-rejected traces were averaged to produce a single trace $I_{CS}^{(i)}(x)$, which was fit once more with the same five-parameter model to extract the per-dataset values of $(I_0, C_0, V_0, I_1, \theta)$.

\begin{figure}[H]
    \centering
    \includegraphics[width=0.4\columnwidth]{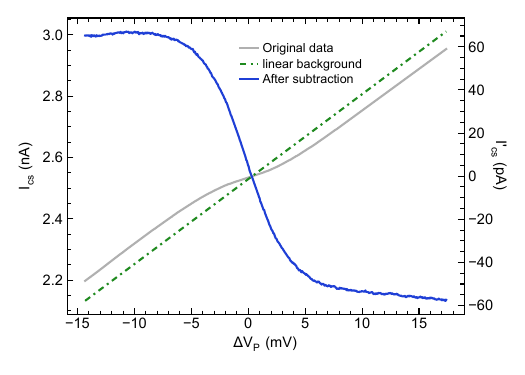}
    \caption{\label{fig:ct_backgroundsub}
    Background subtraction for a representative dataset ($T_\mathrm{mc} = 602\,\mathrm{mK}$).  Grey: dataset-level charge-sensor signal $I_{CS}^{(i)}(x)$ (left axis).  Green dot--dashed: the linear background $C_0\,x + I_0$ removed from each trace.  Blue: the resulting background-subtracted trace $\di^{(i)}(x)$ (right axis), the input to the NRG fit.  The residual sign-changing nature of the slopes on left and right side of the transition in the blue trace motivates the slope correction $C_1$ in the NRG model, further discussed in section \ref{sec:QI}.
    }
\end{figure}

Fitting is more robust if the cross-capacitance between the plunger gate and the charge sensor is first removed, to first order, and the charge-sensor current is offset to lie around zero. After the trace averaging described above, we therefore subtracted a straight line from each averaged trace, chosen so that the wings of the charge transition are approximately flat. The resulting background-subtracted traces $\di^{(i)}(x)$ are the input to the joint NRG fit (Sec.~\ref{sec:nrgproc}); an example is shown in Fig.~\ref{fig:ct_backgroundsub}.

\subsection{NRG fitting procedure}\label{sec:nrgproc}

\subsubsection{NRG library and dimensionless rescaling}\label{sec:nrglib}

NRG calculations of occupation and conductance vs dot energy $\varepsilon$  were precomputed at a single fixed temperature $T_0$ with $\Gamma$ scanned over a discrete grid, producing a family of dimensionless occupation lineshapes $N_\mathrm{NRG}(\varepsilon/\Gamma)$ and conductance curves $G_\mathrm{NRG}(\varepsilon/\Gamma)$, each labelled by the ratio $\Gamma/T_0$.  Because the bandwidth in the NRG calculations is so much larger than $\Gamma$ and $T$, these curves depend only on the ratio $\Gamma/T$ to a very high approximation.  The library generated at $T_0$ can be applied to any $\Gamma$, at any dataset temperature $T_i$, by linearly interpolating between the two library curves whose ratios bracket the target $\Gamma/T_i$. This framework is common to both the entropy and conductance fits; the differences lie in which output quantity is fitted and in the per-dataset parameterisation, as described in the respective sections.

\subsubsection{Joint NRG fit to the occupation lineshape for $\alpha$ and $\Gamma$}\label{sec:jointfit}

\begin{figure}[h]
    \centering
    \includegraphics[width=0.5\columnwidth]{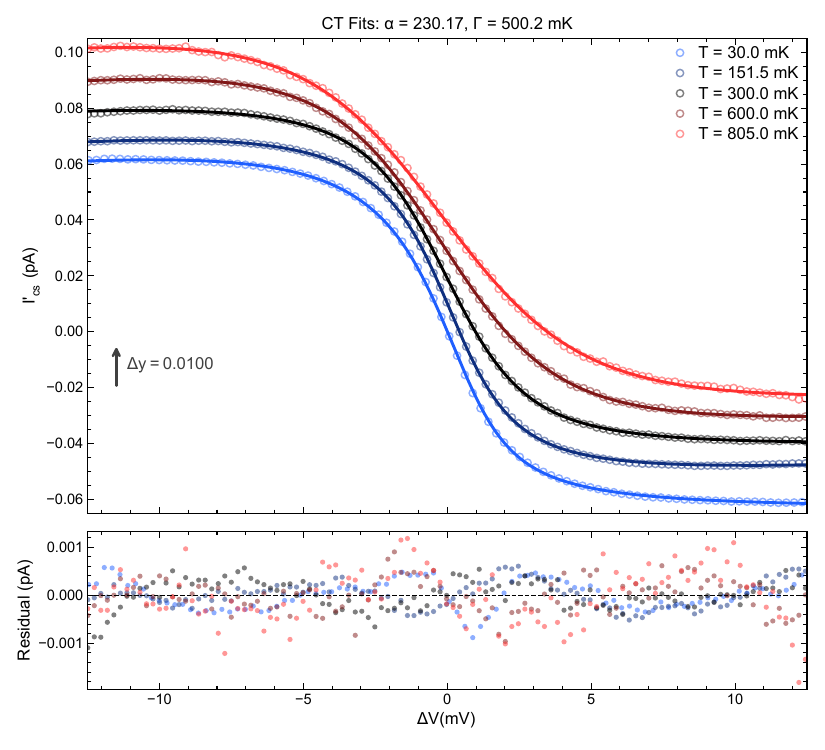}
    \caption{\label{fig:jointnrg}
    Joint NRG fits to the background-subtracted charge-transition traces. Five representative datasets spanning the full $T_\mathrm{mc}$ range are shown (the joint fit uses every dataset). Top: background-subtracted charge-sensor signal (open circles) and NRG model (solid lines) at $T_\mathrm{res} \approx 25$, 150, 300, 600, 800~mK; successive traces are offset vertically by $\Delta y = 0.01\,\mathrm{nA}$ for clarity, with colour running from blue (coldest) to red (hottest). The converged global parameters are $\alpha \approx 230\,\mathrm{mK/mV}$ and $\Gamma \approx 500\,\mathrm{mK}$. Bottom: residuals (data minus model) for the same five datasets.
    }
\end{figure}

In order to determine $\Gamma$ for a given device setting, we use a global fit to the temperature dependence of the charge transition across a broad range of temperatures,
taking advantage of the fact that the hybridization $\Gamma$ and lever-arm $\alpha$ are device parameters that should not depend on temperature.  They could therefore be treated as global parameters shared across the entire temperature series.
For the data in Fig.~\ref{fig:dndtvsn}, the temperature-dependent data includes 10 datasets at $T_\mathrm{res} \approx 25$, 40, 70, 100, 150, 200, 300, 400, 600, 800~mK.  The global fit to these datasets (illustrated in Fig.~\ref{fig:jointnrg}) is described below.

In addition to $\Gamma$ and $\alpha$, each trace also has five per-dataset parameters, referred to collectively as $\{\phi_i\}$: the charge-transition midpoint $V_0$; a linear background described by an offset $I_0$ and slope $C_0$; and the step itself described by an amplitude $I_1$ and an occupation-dependent correction $C_1$ to the slope.  The need for $C_1$ is discussed in more detail in Sec.~\ref{sec:QI}.  The full model for each trace is
\begin{equation}
\di^\mathrm{model}(x) = I_0 + C_0\,x  + \bigl[I_1 + C_1\,x \bigr]\,N_\mathrm{NRG}\!\left(\frac{\alpha\,(x - V_0)}{\Gamma}\right).
\end{equation}

\noindent where $\di^\mathrm{model}(x)$ is evaluated for each dataset using its own $\phi_i$ and the shared $(\alpha, \Gamma)$.  The best-fit parameters are those that minimize mean squared residual, $E(\Gamma, \alpha, \{\phi_i\}\bigr)$,  across all datasets.

Finding this minimum was challenging because of inter-dependencies among the parameters.  The most extreme example of inter-dependence is the partial degeneracy of parameters $\Gamma$ and $\alpha$: both control the energy scale of the charge-transition.  In order to minimize the effect of various interdependences we conducted a staged fit:  First, approximate values of $\Gamma$ and $\alpha$ are obtained by inspection from the temperature at which charge transitions begin to broaden, and initial guesses for $\{\phi_i\}$ are found from a $tanh$ fit.  Using these as starting parameters, $V_0,I_0, I_1$ are optimized in a fit to NRG.  Then, $\Gamma$ and $\alpha$ are added as free parameters.   Slope corrections $C_0$ and $C_1$ were only introduced once the remaining parameters for the NRG fit were seeded, preventing them from absorbing real curvature in the lineshape.

\subsection{Per-temperature view of the Fig.~2 data}\label{sec:pertemp}

\begin{figure}[H]
    \centering
    \includegraphics[width=0.75\columnwidth]{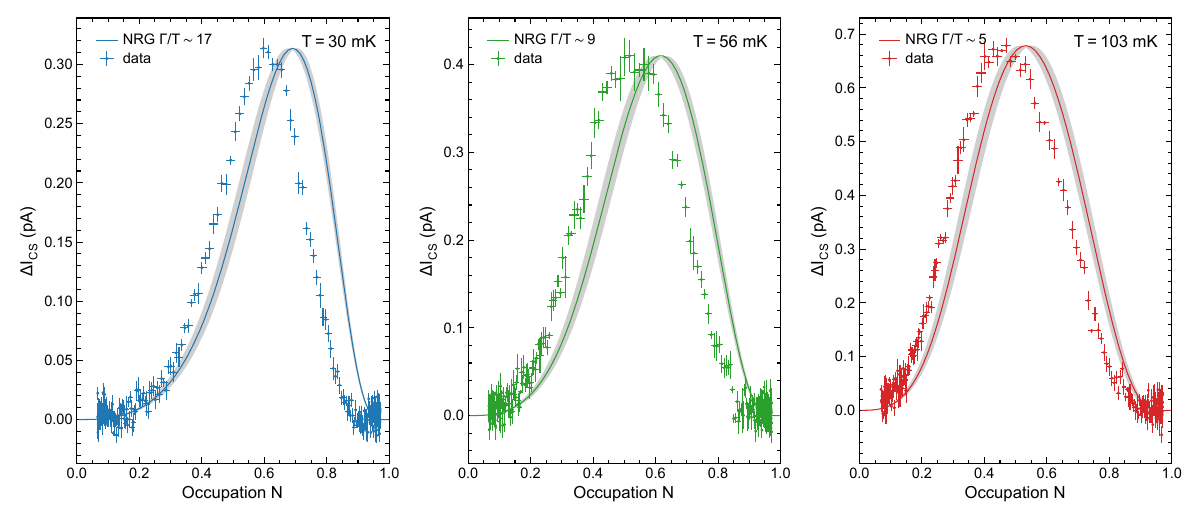}
    \caption{\label{fig:entropyfig_pert}
    The Fig.~2 data shown per temperature. Each panel plots the raw (un-normalized) hot$-$cold charge-sensor difference $\Delta I_{CS}$ across the $0\to1$ transition versus occupation $N$ at one reservoir temperature ($T = 30$, $56$, and $103\,\mathrm{mK}$), together with the corresponding NRG prediction. This is the same data as Fig.~2, presented without normalization and with the temperatures separated for clarity.
    }
\end{figure}

For visual clarity, the data of Fig.~2 are reproduced in Fig.~\ref{fig:entropyfig_pert} with the three temperatures plotted in separate panels and the raw, un-normalized charge-sensor difference $\Delta I_{CS}$ on the vertical axis, rather than the normalized $dN/dT$ of Fig.~2.

\begin{table}[H]
\centering
\caption{Parameters for Fig.~2 (entropy $dN/dT(N)$).}
\label{tab:params_fig2}
\begin{ruledtabular}
\begin{tabular}{@{\extracolsep{\fill}} l l l}
Parameter & Value & Unit \\ \hline
Heater step $dT$\,$^\mathrm{a}$ & \textbf{3}, 5, 10 & mK \\
Charge-sensor bias & 210 & $\mu$V \\
Runs per temperature & 8, 4, 4 & -- \\
Hybridization $\Gamma_\mathrm{best}$\,$^\mathrm{c}$ & 500 & mK \\
$\Gamma$ interval $[\Gamma_\mathrm{lo},\Gamma_\mathrm{hi}]$ & 440--575 & mK \\
Lever arm $\alpha$\,$^\mathrm{b}$ & 230 & mK/mV \\
$\Gamma/T$ (per $T$) & 16.7, 8.9, 4.9 & -- \\
\end{tabular}
\end{ruledtabular}
\begin{flushleft}\footnotesize
$^\mathrm{a}$ $dT=3$~mK is calibrated at $T=30$~mK (bold); at 56 and 103~mK the heating targeted $\Delta T/T\approx0.1$ ($\approx$5 and 10~mK) but was not independently calibrated. \\
$^\mathrm{b}$ manuscript quotes $\alpha\approx231$. \\
$^\mathrm{c}$ All $\Gamma$ values quoted in this work (tables and text) are rounded to the nearest 5~mK, far below the fit uncertainty; the $\Gamma/T$ ratios in the tables are computed from the unrounded values. \\
\end{flushleft}
\end{table}

\section{Conductance and charge-transition analysis (Fig.~3)}\label{sec:condctanalysis}

\noindent\textbf{Overview:} The parametric $G(N)$ comparison in Fig.~3 was assembled from conductance and charge-sensor data acquired simultaneously (Sec.~\ref{sec:condmeas}) at multiple temperatures, then analyzed in several stages.  First, repeated traces collected at each temperature were aligned and averaged (Sec.~\ref{sec:tracealign}).  Second, the conductance traces were fit to the NRG conductance lineshape to determine a conductance estimate of hybridization, $\Gamma_G$ (Sec.~\ref{sec:condctfit}), and a separate fit of the charge-transition data  provided an occupation-based estimate, $\Gamma_N$. The two were averaged, giving $\Gamma_\mathrm{avg}=(\Gamma_G+\Gamma_N)/2$.  Third, each charge-transition dataset was fit to the NRG occupation lineshape, holding $\Gamma$ fixed at $\Gamma_\mathrm{avg}$, to obtain the gate-voltage-to-$N$ mapping.  This converted the simultaneously measured $G(\Delta{V}_P)$ into the $G(N)$ of Fig.~3 (Sec.~\ref{sec:occfit}).  Finally, the uncertainties shown in Fig.~3 were evaluated (Secs.~\ref{sec:gammaerr} and \ref{sec:occerr}): the shaded band reflects the uncertainty in $\Gamma$ (Sec.~\ref{sec:gammaerr}), and the horizontal error bars reflect the uncertainty in the extracting occupation from the charge transition data (Sec.~\ref{sec:occerr}).

\subsection{Trace alignment and averaging}\label{sec:tracealign}

{\raggedright
At each temperature the repeated dot current traces were aligned by shifting each to a common peak position, removing slow gate-voltage drift and jumps due to unstable offset charges, and then averaged.  The same shift was applied to the corresponding charge-sensor trace, keeping $I_{CS}$ and $G$ on a common gate-voltage axis.  Outlier traces were identified and excluded,  the surviving traces were averaged, the series resistance of the measurement lines and ohmic contacts was removed from dot-current measurements, then a linear background was subtracted from $I_{CS}$, giving $I'_{CS}$.
\par}

\subsection{NRG fitting to determine $\Gamma$ and $\alpha$}\label{sec:nrgfit}
{\raggedright
A single device hybridization $\Gamma$ sets the energy axis of both the conductance lineshape and the occupation lineshape. We can therefore estimate $\Gamma$ from either channel. The conductance traces give one estimate, $\Gamma_G$, and the charge-transition traces give an independent one, $\Gamma_N$ (both in Sec.~\ref{sec:condctfit}). Because the two channels carry different backgrounds and noise and constrain the lineshape differently, these estimates differ slightly, and neither is \emph{a priori} more reliable. We therefore use their average, $\Gamma_\mathrm{avg}=(\Gamma_G+\Gamma_N)/2$, as the single value at which the occupation is extracted, and we use the offset between $\Gamma_G$ and $\Gamma_N$---together with the shape of the error curve in each channel---to set the uncertainty band on $\Gamma$ (Sec.~\ref{sec:gammaerr}).
\par}

\subsubsection{Determination of $\Gamma_G$ and $\Gamma_N$}\label{sec:condctfit}

The conductance estimate $\Gamma_G$ was extracted from a simultaneous fit of the $G(\Delta{V}_P)$ traces to NRG at several temperatures, with $\alpha$ and $\Gamma$ shared across temperatures as in the occupation fit of Sec.~\ref{sec:nrgproc}. The conductance fit was based on the same NRG library as in Sec.~\ref{sec:nrgproc}, using the dimensionless conductance $G_\mathrm{NRG}(\varepsilon/\Gamma)$ in place of the occupation $N_\mathrm{NRG}(\varepsilon/\Gamma)$. Unlike the charge-sensor signal, the conductance data carries no cross-capacitance background, so the per-dataset model included only an amplitude $A^{(i)}$ and a gate-voltage offset $V_0^{(i)}$, with no background slope or vertical offset.
We then scan $\Gamma$ on a grid in order to estimate the uncertainty in $\Gamma_G$ (Sec.~\ref{sec:gammaerr}).

The occupation estimate $\Gamma_N$ is obtained in the same way as the $\Gamma$ extraction described in Sec.~\ref{sec:entanalysis}.

\subsection{Extraction of occupation}\label{sec:occfit}

\begin{figure}[H]
    \centering
    \includegraphics[width=0.6\columnwidth]{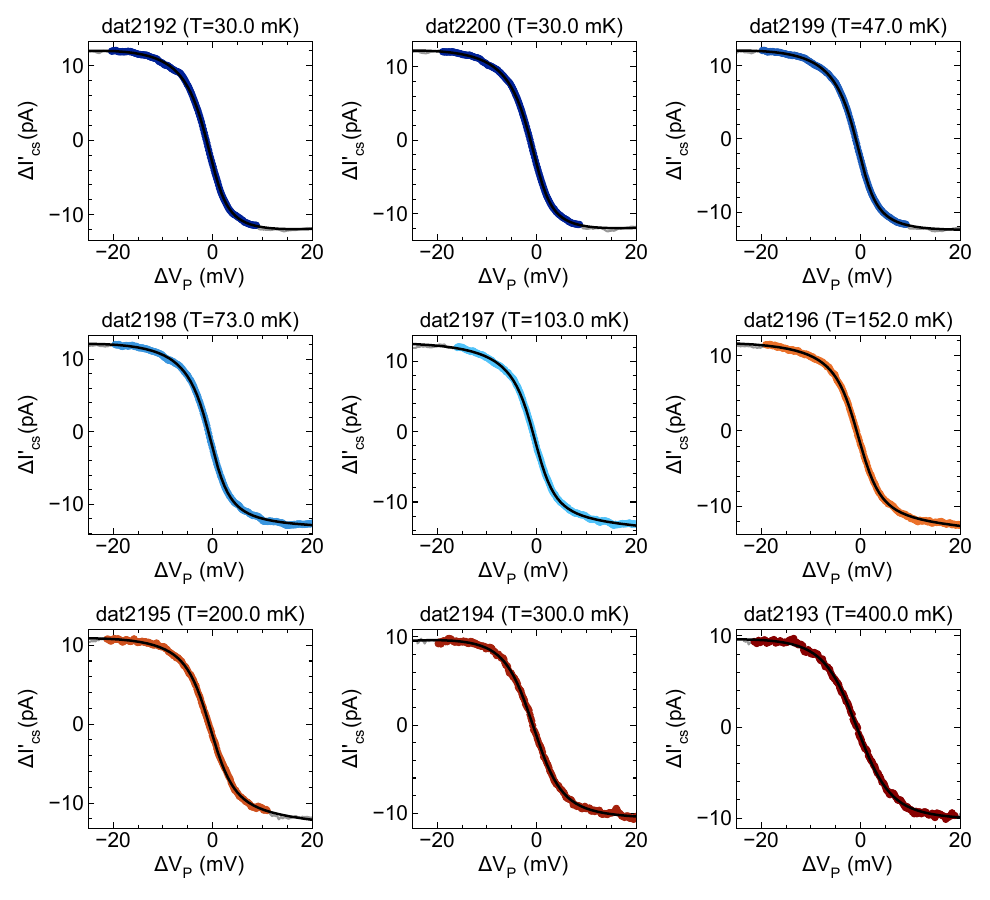}
    \caption{\label{fig:ctfitting}
    Individual charge-transition fits for the device of Fig.~3, one panel per temperature ($T = 30$--$400$~mK). Coloured points: background-subtracted charge-sensor signal $I'_{CS}$ versus gate voltage; black curves: the NRG occupation model of Sec.~\ref{sec:occfit} (global $\alpha$ and $\Gamma$, with the per-dataset amplitude, vertical offset, and slope corrections). These fits define the gate-voltage-to-$N$ mapping $N(\Delta{V}_P)$ used to form $G(N)$ in Fig.~3.
    }
\end{figure}

{\raggedright
As in Sec.~\ref{sec:ndet}, the gate voltage-to-$N$ mapping for Fig.~\ref{fig:gvsn} was determined by fitting $I'_{CS}(\Delta\vp)$ to NRG occupation curves, with $\Gamma$ held fixed at $\Gamma_\mathrm{avg}$.  The individual charge-transition fits across the temperature series are shown in Fig.~\ref{fig:ctfitting}.

\par}

\begin{table}[t]
\centering
\caption{Parameters for Fig.~3 (conductance $G(N)$).}
\label{tab:params_fig3}
\begin{ruledtabular}
\begin{tabular}{@{\extracolsep{\fill}} l l l}
Parameter & Value & Unit \\ \hline
Charge-sensor bias & 50 & $\mu$V \\
AC excitation $V_\mathrm{bias}$ & 0.911 & $\mu$V \\
Series resistance $R_s$ & 21.2 & k$\Omega$ \\
Hybridization $\Gamma_G$ (conductance basin) & 400 & mK \\
Lever arm $\alpha_G$ (conductance) & 431 & mK/mV \\
Hybridization $\Gamma_N$ (CT basin) & 300 & mK \\
Lever arm $\alpha_N$ (CT) & 415 & mK/mV \\
Combined $\Gamma$ band $[\Gamma_\mathrm{lo},\Gamma_\mathrm{hi}]$\,$^\mathrm{a}$ & 300--400 & mK \\
Hybridization $\Gamma$ (averaged, used in Fig.~3) & 350 & mK \\
Lever arm $\alpha$ (used in Fig.~3) & 418 & mK/mV \\
$\Gamma/T$ (for each $T$) & 11.6, 3.4, 0.9 & -- \\
\end{tabular}
\end{ruledtabular}
\begin{flushleft}\footnotesize
$^\mathrm{a}$ $[\Gamma_N,\Gamma_G]$: CT fit to conductance fit. \\
\end{flushleft}
\end{table}

\newpage
\section{Uncertainty analysis}\label{sec:uncertainty}

\subsection{Determination of $\Gamma$ (entropy measurement)}\label{sec:gammadet}

\begin{figure*}[h]
    \centering
    \sbox\panelB{\includegraphics[height=5.2cm]{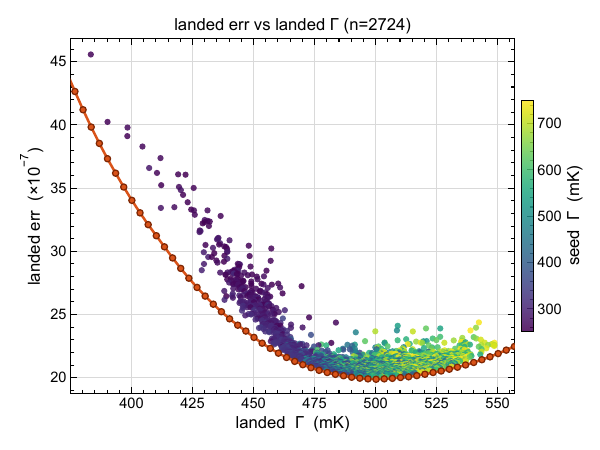}}%
    \sbox\panelA{\includegraphics[height=5.2cm]{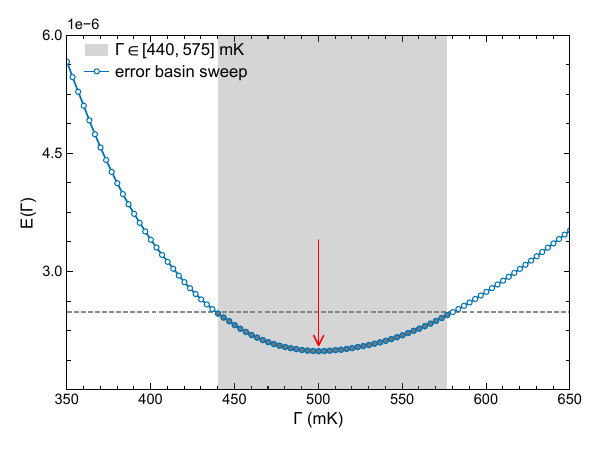}}%
    \begin{minipage}[b]{\wd\panelB}
        \textbf{(a)}\par\smallskip
        \usebox\panelB
    \end{minipage}\hfill
    \begin{minipage}[b]{\wd\panelA}
        \textbf{(b)}\par\smallskip
        \usebox\panelA
    \end{minipage}
    \caption{\label{fig:gammabasin_ent}
     \textbf{(a)} Robustness of the global NRG fit to charge transition data to varying initialization, for the device setting in Fig.~\ref{fig:dndtvsn}. Each point is one free fit (over $\Gamma$, $\alpha$, $\{\phi_i\}$) from a seed $\Gamma_\mathrm{seed}$ (colour), plotted at its landed $\Gamma$ (horizontal) and final fit error (vertical). Independent of the seed, runs land near $\Gamma\approx450$--$550\,\mathrm{mK}$; their lower envelope coincides with $E(\Gamma)$ (orange) in (b).
     \textbf{(b)} Constrained error curve $E(\Gamma)$ (5~mK grid) for the device setting in Fig.~\ref{fig:dndtvsn}. Shading: the interval $[\Gamma_\mathrm{lo}, \Gamma_\mathrm{hi}]$ where $E(\Gamma) < 1.25\,E_\mathrm{min}$ (dashed line: threshold); arrow: best fit, $\Gamma\approx500\,\mathrm{mK}$.
    }
\end{figure*}

The hybridization $\Gamma$ for the entropy measurement is set by the joint NRG fit to the charge-transition series (Sec.~\ref{sec:jointfit}). Because $\Gamma$ is nearly degenerate with $\alpha$ and interacts with the per-dataset parameters, the fit error has many local minima, and its global minimum is shallow in $\Gamma$. To locate that minimum reliably we repeated the fit from many random initializations. Independent of the seed, the runs collapse onto a single shallow valley: seeds spanning $\Gamma_\mathrm{seed}=250$--$750\,\mathrm{mK}$ land with $\Gamma$ clustered at $450$--$550\,\mathrm{mK}$, and the fit error bottoms out near $\Gamma_\mathrm{best}\approx500\,\mathrm{mK}$ [Fig.~\ref{fig:gammabasin_ent}(a), for the setting of Fig.~\ref{fig:dndtvsn}].

The difficulty in pinning down $\Gamma$ is compounded by the shallowness of the minimum itself. We characterize the fit landscape along $\Gamma$ with a \emph{constrained error curve} $E(\Gamma)$: at each point of a grid, $\Gamma$ is held fixed while the fit is re-minimized over $\alpha$ and all per-dataset parameters $\{\phi_i\}$, and $E(\Gamma)$ is the resulting best (minimized) error [Fig.~\ref{fig:gammabasin_ent}(b)]. Its global minimum defines $\Gamma_\mathrm{best}$, its steepness measures how strongly the data constrain $\Gamma$, and it coincides with the lower envelope of the random-seed scatter of panel (a). For the NRG fits to the charge-transition data, $E(\Gamma)$ is strikingly shallow, varying only weakly over a wide range of $\Gamma$; converting this into an uncertainty is deferred to Sec.~\ref{sec:gammaunc}. The same construction, applied to the occupation, gives the uncertainty in $N$ (Sec.~\ref{sec:ndet}).

\subsection{Determination of $\Gamma$ and its uncertainty (conductance measurement)}\label{sec:gammaerr}

In the conductance measurement, $\Gamma$ sets the energy scale of both the conductance peak and the charge transition, giving two independent estimates: $\Gamma_G$ from the conductance fit and $\Gamma_N$ from the charge-transition fit (Sec.~\ref{sec:condctfit}). The conductance fit is much better conditioned---it converged directly to its minimum, without the seed ensemble the charge-transition fit required (Sec.~\ref{sec:gammadet})---though not necessarily more accurate, since the two channels carry different backgrounds and noise. For the device tuning in Fig.~\ref{fig:gvsn}, the global fit to the conductance temperature series yields $\Gamma_G=400\,\mathrm{mK}$, versus $\Gamma_N=300\,\mathrm{mK}$ from the simultaneously measured charge transitions. With no strong reason to trust one over the other, we take their average, $\Gamma_\mathrm{avg}=350\,\mathrm{mK}$, as the value used in Fig.~3, and the interval between them, $[\Gamma_N,\Gamma_G]$, as the uncertainty; this sets the shaded NRG bands in Fig.~3. The same procedure on the second device gives $\Gamma_G=415$ and $\Gamma_N=305\,\mathrm{mK}$ (Sec.~\ref{sec:dev2_gn}).

\begin{figure}[H]
    \centering
    \includegraphics[width=0.5\columnwidth]{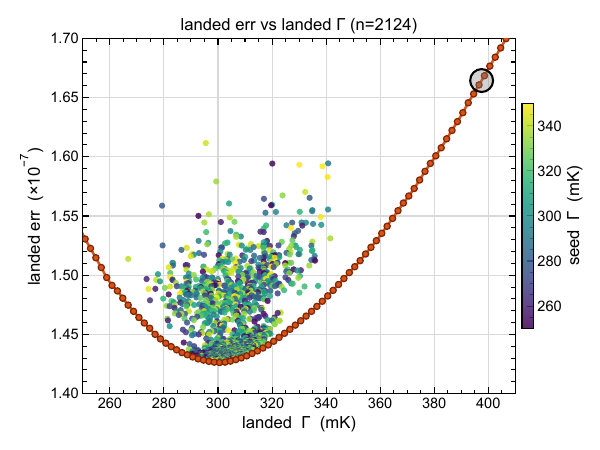}
    \caption{\label{fig:gammabasin_cond}
    Robustness of the global NRG fit to charge transition data to varying initialization, for the device setting in Fig.~\ref{fig:gvsn}, together with the constrained fit error, analogous to Fig.~\ref{fig:gammabasin_ent}.  The circled datapoint shows the increase in fit error when $\Gamma$ is fixed to $\Gamma_G$ during the fit to the charge-transition data.
    }
\end{figure}

\subsection{Uncertainty in $\Gamma$ (entropy measurement)}\label{sec:gammaunc}

For the entropy measurement only the charge-transition channel is available, so the shallow $E(\Gamma)$ of Fig.~\ref{fig:gammabasin_ent}(b) is all we have. Because $E$ is not normalized to an independent noise estimate, the uncertainty in $\Gamma$ cannot come from the curvature of the minimum or from a $\Delta\chi^2$ criterion; instead, we use the $\Gamma_N$--$\Gamma_G$ contrast described in Sec.~\ref{sec:gammaerr} to gauge how large a fit-error increase corresponds to a still-plausible $\Gamma$. Figure~\ref{fig:gammabasin_cond} shows the constrained charge-transition fit error $E(\Gamma)$ for Fig.~\ref{fig:gvsn}, together with the random-seed fit results. Moving from the charge-transition minimum $\Gamma_N=300\,\mathrm{mK}$ to the conductance value $\Gamma_G=400\,\mathrm{mK}$ raises the fit error by $16\%$ for Fig.~3, and by $30\%$ for the equivalent measurement in a second device (see Sec.~\ref{sec:dev2_gn}). Taking a characteristic $25\%$ from this range, we set the interval $[\Gamma_\mathrm{lo},\Gamma_\mathrm{hi}]$ by $E(\Gamma)<1.25\,E_\mathrm{min}$. For Fig.~\ref{fig:gammabasin_ent}, this gives the asymmetric interval $440\,\mathrm{mK}\lesssim\Gamma\lesssim575\,\mathrm{mK}$, which sets the shaded NRG bands in Fig.~2(b).

\subsection{Uncertainty in $N$}\label{sec:ndet}

The occupation $N$ is obtained by mapping gate voltage through the NRG lineshape at fixed $(\alpha,\Gamma)$, and is most sensitive to the transition midpoint $V_0$ (where $N=1/2$), in interplay with the two slope corrections that can tilt the lineshape to partly compensate an incorrect $V_0$. We bound it as we did $\Gamma$, using the constrained error curve $E(V_0)$ (defined as for $E(\Gamma)$ in Sec.~\ref{sec:gammadet}): with $\alpha$ and $\Gamma$ fixed, $V_0$ is stepped on a grid, the remaining per-dataset parameters re-minimized at each step, and the interval taken as the range of $V_0$ over which $E(V_0)$ stays within a fractional tolerance---set by inspecting where the slopes can no longer compensate and the fit visibly departs from the data. This was done separately for each dataset. Mapping the two bounds to occupation brackets $N$ at each gate voltage, and the spread is the horizontal error bar in Figs.~2 and 3. \label{sec:occerr} In Fig.~3 only these horizontal bars appear: the vertical uncertainty in $G$ is set by measurement noise and is smaller than the marker.
\newpage
\section{Reproduction of Fig.~2 on a second device}\label{sec:dev2_dndt}

\begin{figure}[h]
    \centering
    \includegraphics[width=0.65\columnwidth]{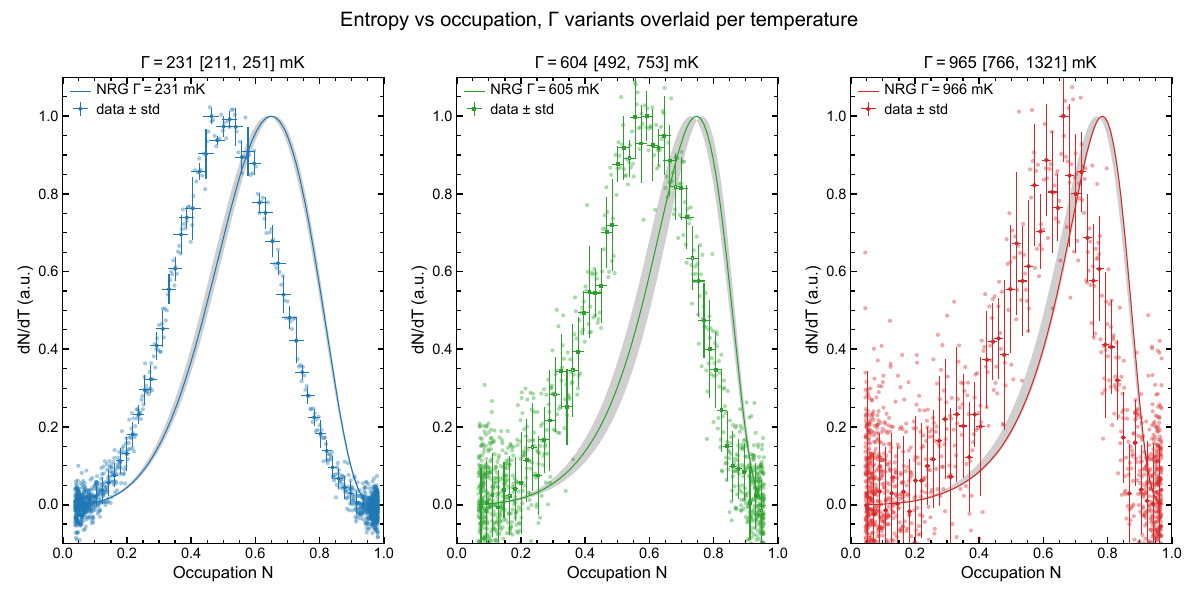}
    \caption{\label{fig:dev2_dndtvsn}
    Entropy signature $dN/dT(N)$ on a second device, measured at $20\,\mathrm{mK}$ for three values $\Gamma:$ 230, 605 and 965 mK (left to right; bracketed ranges give the $\Gamma$ uncertainty). Markers: measured $dN/dT$ ($\pm$ std), with faint points showing the individual runs; solid curves: NRG at the stated $\Gamma$ with a shaded $\Gamma$-uncertainty band. As $\Gamma$ increases the peak shifts further toward $N>1/2$, the signature of stronger Kondo screening; each panel is normalized to its peak. This complements Fig.~2, which instead varied temperature at fixed $\Gamma$.
    }
\end{figure}

The entropy signature of Fig.~2---the asymmetry of $dN/dT(N)$   toward $N>1/2$---was reproduced on a second device. Rather than tracking $dN/dT(N)$ at several temperatures for a single $\Gamma$, as in Fig.~2, here the temperature was held at $20\,\mathrm{mK}$ and $\Gamma$ was varied for three tunings of the dot. As $\Gamma$ increases, $\tk$ rises and the local moment stays Kondo-screened further onto the $N>1/2$ side of the transition, so the peak of $dN/dT(N)$ shifts further toward $N>1/2$ (Fig.~\ref{fig:dev2_dndtvsn}). NRG calculations with no free parameters reproduce this trend, mirroring the temperature dependence of Fig.~2, but also the consistent shift to lower $N$ of data versus NRG.

\begin{table}[h]
\centering
\caption{Parameters for Fig.~\ref{fig:dev2_dndtvsn} (entropy $dN/dT(N)$ reproduced on a second device at three hybridizations, $T=20$~mK).}
\label{tab:params_dev2}
\begin{ruledtabular}
\begin{tabular}{@{\extracolsep{\fill}} l c c c}
 & Low $\Gamma$ & Medium $\Gamma$ & High $\Gamma$ \\ \hline
Hybridization $\Gamma$ (mK) & 230 & 605 & 965 \\
Lever arm $\alpha$ (mK/mV) & 60.0 & 208 & 371 \\
$\Gamma$ interval $[\Gamma_\mathrm{lo},\Gamma_\mathrm{hi}]$ (mK) & 210--250 & 490--755 & 765--1320 \\
$\Gamma/T$ & 11.6 & 30.2 & 48.3 \\
\end{tabular}
\end{ruledtabular}
\end{table}

\newpage
\section{Reproduction of Fig.~3 on a second device}\label{sec:dev2_gn}

The Kondo signature of Fig.~3---the shift of the $G(N)$ peak toward $N>1/2$---was reproduced on a second device of the same design.

\begin{figure}[h]
    \centering
    \includegraphics[width=0.5\columnwidth]{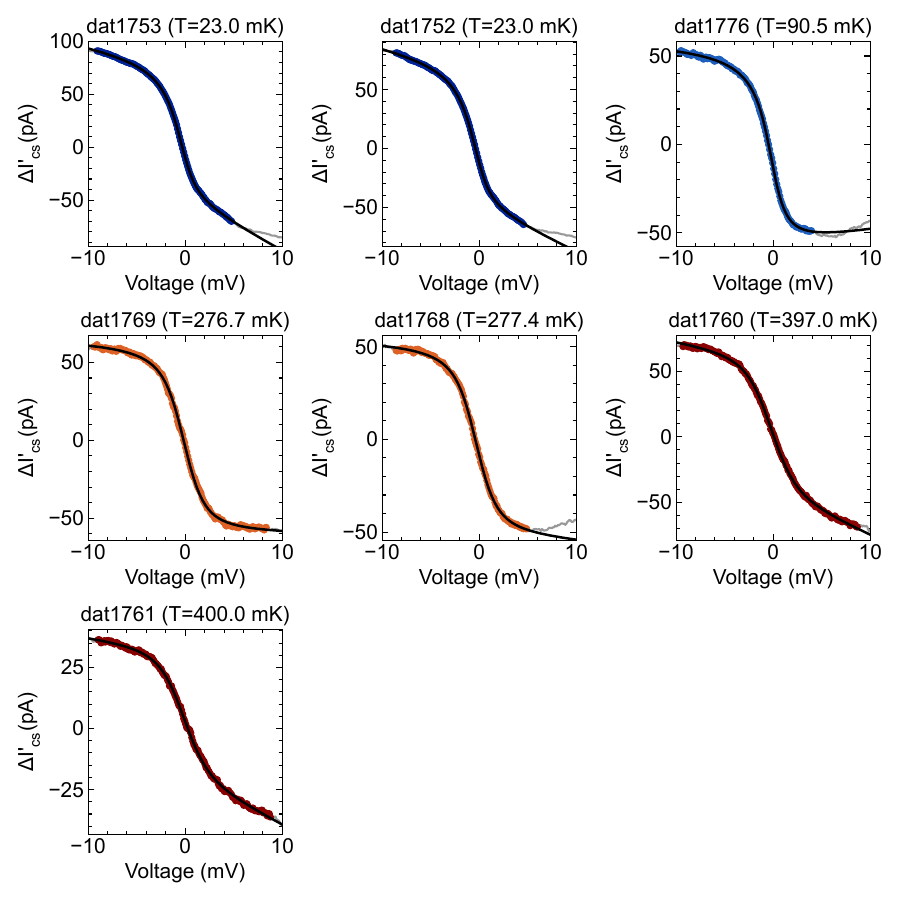}
    \caption{\label{fig:dev2_condct}
    Individual charge-transition fits for a second device, one panel per temperature ($T = 23$--$400$~mK). Coloured points: background-subtracted charge-sensor signal; black curves: NRG occupation model with linear background. These fits provide the occupation $N$ used in Fig.~\ref{fig:dev2_gvsn}.
    }
\end{figure}

\begin{figure}[h]
    \centering
    \includegraphics[width=0.6\columnwidth]{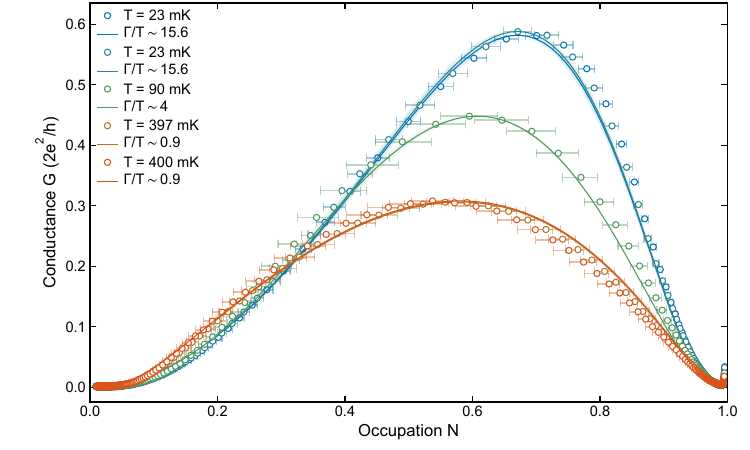}
    \caption{\label{fig:dev2_gvsn}
    Conductance $G$ versus occupation $N$ for a second device at three temperatures, analyzed as in Fig.~3. Markers: data with horizontal (occupation) error bars; solid curves: NRG at the best-fit $\Gamma$; shaded bands: NRG over the $\Gamma$ uncertainty interval. 
    }
\end{figure}

\begin{table}[t]
\centering
\caption{Parameters for the conductance $G(N)$ reproduced on a second device (Fig.~\ref{fig:dev2_gvsn}).}
\label{tab:params_dev2_cond}
\begin{ruledtabular}
\begin{tabular}{@{\extracolsep{\fill}} l l l}
Parameter & Value & Unit \\ \hline
Temperatures & 23, 90, 400 & mK \\
Charge-sensor bias & 100 & $\mu$V \\
AC excitation $V_\mathrm{bias}$ & 1 & $\mu$V \\
Series resistance $R_s$ & 22 & k$\Omega$ \\
Hybridization $\Gamma_G$ (conductance basin) & 415 & mK \\
Lever arm $\alpha_G$ (conductance) & 163 & mK/mV \\
Hybridization $\Gamma_N$ (CT basin) & 305 & mK \\
Lever arm $\alpha_N$ (CT) & 165 & mK/mV \\
Combined $\Gamma$ band $[\Gamma_\mathrm{lo},\Gamma_\mathrm{hi}]$\,$^\mathrm{a}$ & 305--415 & mK \\
Hybridization $\Gamma$ (averaged, used in Fig.~\ref{fig:dev2_gvsn}) & 360 & mK \\
Lever arm $\alpha$ (used in Fig.~\ref{fig:dev2_gvsn}) & 156 & mK/mV \\
$\Gamma/T$ (per $T$) & 15.6, 4, 0.9 & -- \\
Conductance-peak location $N_\mathrm{peak}$ & 0.674, 0.605, 0.575 & -- \\
\end{tabular}
\end{ruledtabular}
\begin{flushleft}\footnotesize
$^\mathrm{a}$ $[\Gamma_N,\Gamma_G]$: CT fit to conductance fit. \\
\end{flushleft}
\end{table}

\section{Quantum Interference Contribution to Charge Sensor Traces}\label{sec:QI}

A small but non-zero capacitive coupling between the dot plunger gate and the charge-sensor reservoirs caused the plunger sweep to modulate the conductance of the sensor through mesoscopic interference.  Because this coupling was weak, the correction appeared not as overt conductance fluctuations but as a smooth, gate-dependent distortion of the charge-transition background; complexity in that background made fits to NRG much less robust, especially because it added an effective difference in background slope between the two sides of the transition.

\begin{figure}[h]
    \centering
    \includegraphics[width=0.5\columnwidth]{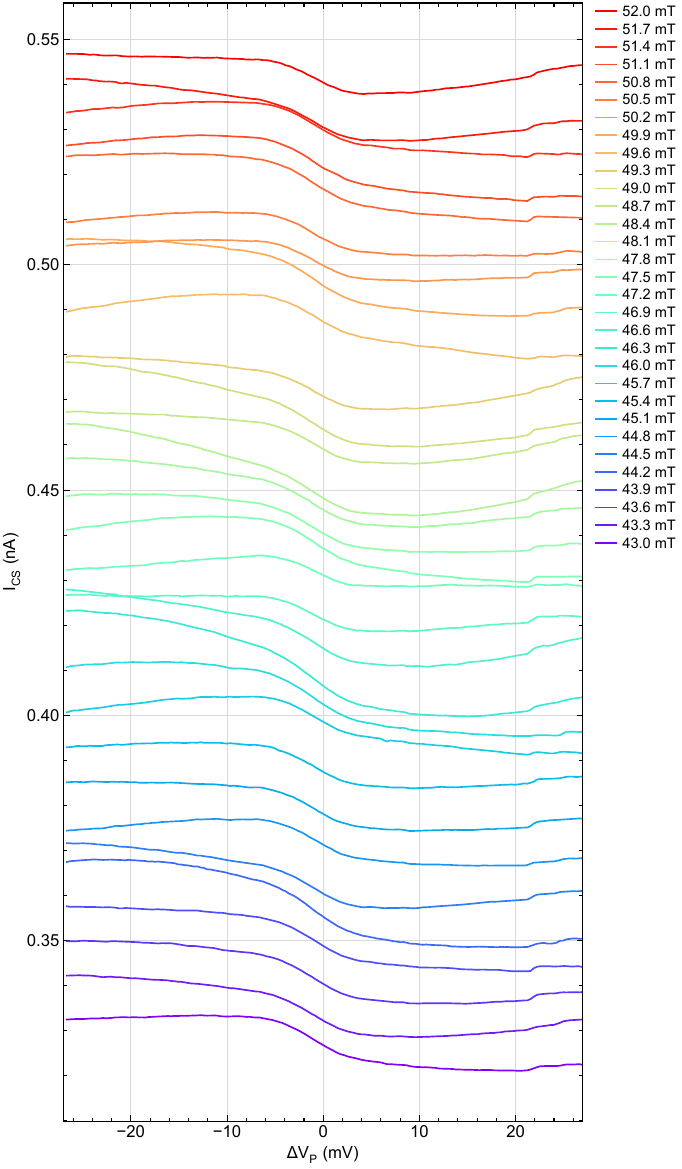}
    \caption{\label{fig:bperp_averaging}
    Averaged charge-sensor traces across the $0\to1$ charge transition, measured at perpendicular magnetic fields stepped from $-43$~mT (bottom, violet) to $-52$~mT (top, red) in $-0.3$~mT steps.  Traces are aligned at the transition midpoint and offset vertically by $7$~pA for clarity.
    }
\end{figure}

Two observations identified the distortion as quantum interference.  First, it rearranged completely under minute changes in $B_\perp$, of order the field that threads one flux quantum through a phase-coherent area---sub-millitesla for our high-mobility two-dimensional electron gas at tens of millikelvin.  Second, it was strongly suppressed at elevated temperature (at and above $\sim100$~mK) or at larger charge-sensor bias.  Elevated temperatures suppress the phase coherence responsible for this effect, while larger sensor bias effectively performs averaging over a larger energy window.
Figure~\ref{fig:bperp_averaging} displays the effect under conditions chosen to exaggerate it: a charge-sensor bias of only $25\,\mu\mathrm{V}$ and an electron temperature of $30$~mK.  The $0\to1$ transition is shown for $B_\perp$ stepped from $43$ to $52$~mT in $0.3$~mT increments: the distortion took an essentially independent form at each step, while the underlying charge transition was unchanged.  Averaging each trace over a range of $B_\perp$---the first step in the trace processing for the entropy measurement---therefore averaged the correction away while preserving the transition.  Although much smaller at the higher bias used for the entropy data, the distortion was nevertheless a primary source of uncertainty in our experiment, so this averaging was applied throughout.

An important consequence of the quantum interference was the need to include a correction $C_1$, on the occupied side of the charge transition, to the overall slope $C_0$.  Although the correction became less significant when measuring at higher $T$ and $V_{CS}$, and by averaging over a greater range of $B_\perp$, it was always included as a fit parameter.

\newpage
\section{Differential conductance maps confirming Kondo character}\label{sec:condmaps}

\begin{figure}[h]
    \centering
    \includegraphics[width=0.9\columnwidth]{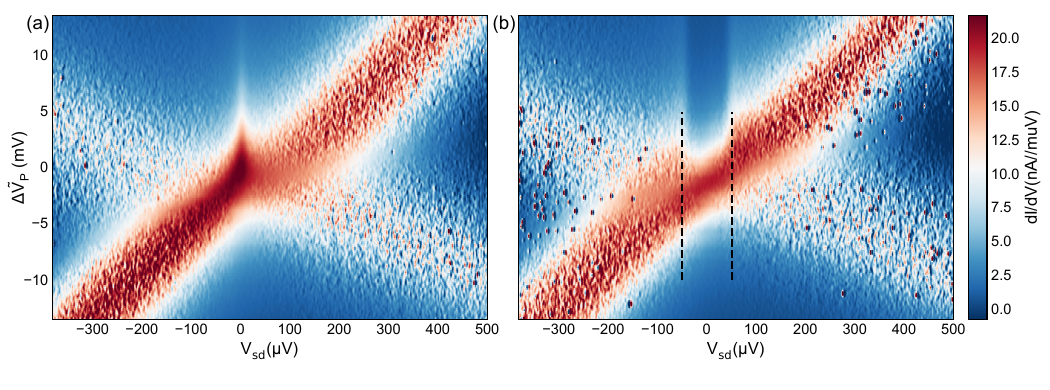}
    \caption{\label{fig:supp_didv}
    Differential conductance maps at $B_\parallel = 0$ and $2\,\mathrm{T}$. Colour-scale plots of $\mathrm{d}I/\mathrm{d}V$ as a function of source-drain bias $V_\mathrm{sd}$ (horizontal axis) and virtual plunger gate detuning $\Delta{V}_P$ (mV, vertical axis), recorded at the device tuning used for all measurements in Fig.~3. (a)~At $B_\parallel = 0\,\mathrm{mT}$, a zero-bias anomaly (ZBA) is visible in the 0--1 charge valley as a sharp ridge of enhanced conductance at $V_\mathrm{sd} = 0$. (b)~At $B_\parallel = 2\,\mathrm{T}$ the ZBA splits into two peaks displaced symmetrically about zero bias. Dashed lines mark the expected peak positions for the bulk GaAs $g$-factor $g = -0.44$; the observed splitting is marginally smaller. The field-induced splitting confirms the Kondo origin of the ZBA.
    }
\end{figure}

Figure~\ref{fig:supp_didv} shows differential conductance $\mathrm{d}I/\mathrm{d}V$ measured as a function of source-drain bias $V_\mathrm{sd}$ and virtual plunger gate detuning $\Delta{V}_P$, providing the device characterisation underlying the data presented in Fig.~3 of the main text. These maps were acquired at the same device tuning as the conductance measurement of Fig.~3; $\Delta{V}_P$ is measured in mV relative to the charge-degeneracy point and is directly comparable to the gate-voltage axis of Fig.~3.

At zero magnetic field [panel (a)], a pronounced zero-bias anomaly (ZBA) is visible in the 0--1 charge valley as a sharp ridge of enhanced conductance centred at $V_\mathrm{sd} = 0$. The ZBA persists across the valley and is absent in the adjacent even-occupation valleys, consistent with spin-$\tfrac{1}{2}$ Kondo screening. At $B_\parallel = 2\,\mathrm{T}$ [panel (b)], the ridge splits into two peaks displaced symmetrically about zero bias. The dashed lines mark the positions expected for the bulk GaAs $g$-factor $g = -0.44$, giving a Zeeman splitting $|g|\mu_B B_\parallel \approx 50\,\mu\mathrm{eV}$; the observed splitting is marginally smaller. This field-induced splitting is a hallmark signature of the Kondo effect and confirms that the ZBA is not a trivial resonance. Together, the two panels establish the Kondo character of the device at the operating point used for the measurements in Fig.~3.

\end{document}